\documentclass[aps,pre,reprint,superscriptaddress,raggedbottom,floatfix]{revtex4-2}
\usepackage{amsmath,amssymb,amsthm,booktabs,pgfplots,etoolbox}
\pgfplotsset{compat=1.18}
\definecolor{tiafH}{HTML}{D55E00}
\definecolor{tiafV}{HTML}{0072B2}
\definecolor{tiafD}{HTML}{009E73}
\usepackage[hidelinks]{hyperref}
\hypersetup{pdfauthor={Weiqi Jiang and Pengcheng Hou},pdftitle={Weak-coupling selection and finite-torus entropy in the triangular Ising antiferromagnet}}
\newcommand{\Z}{\mathbb Z}
\newcommand{\dtot}{D_{\mathrm{tot}}}
\newcommand{\dd}{\mathrm d}
\newcommand{\Cl}{\operatorname{Cl}_2}
\newcommand{\sD}{s_{\mathrm d}}
\newcommand{\sS}{s_{\mathrm s}}
\newtheoremstyle{aps}{\topsep}{\topsep}{\normalfont}{\parindent}
  {\itshape}{.}{0.5em}{}
\theoremstyle{aps}

\expandafter\patchcmd\csname\string\proof\endcsname{\hskip\labelsep}{\hskip\labelsep\hskip\parindent}{}
  {\PackageError{main_pre}{Cannot patch proof indentation}
    {Check the amsthm proof definition.}}
\newtheorem{theorem}{Theorem}
\newtheorem{proposition}{Proposition}
\newtheorem{lemma}{Lemma}
\begin{document}
\title{Weak-coupling selection and finite-torus entropy
in the triangular Ising antiferromagnet}
\author{Weiqi Jiang}
\affiliation{Institute of Theoretical Physics, Chinese Academy of Sciences, Beijing 100190, China}
\author{Pengcheng Hou}
\email[Corresponding author: ]{houpc@hfnl.cn}
\affiliation{Hefei National Laboratory, Hefei 230088, China}
\date{October 28, 2026}
\begin{abstract}
An antiferromagnetic second-neighbor interaction favors stripe order in the
triangular-lattice Ising antiferromagnet, but uniform stripes cannot close
when both primitive periods are odd. At each fixed odd size $L\geq5$, we minimize this
interaction over the ground states of the zero-field nearest-neighbor model
on an $L\times L$ torus. We prove a minimum total interface length of $4L$
and classify all $L^2(L^2-1)/2$ selected states. Under the uniform measure
on these states, the area fractions of the three stripe orientations
approach a uniform distribution over all nonnegative triples summing to one.
Thus selection reduces an exponential degeneracy to polynomial growth,
while the whole-sample orientation fractions remain broadly distributed.
We also derive explicit
two-sided entropy-density bounds for periodic spin ground states and
honeycomb dimer coverings on $M\times N$ tori, where $M,N$ are the periods
along the two primitive lattice directions. For all $M,N\geq3$, the errors
relative to the bulk residual entropy are bounded by explicit constants
times $\min(M,N)^{-2}$, without an aspect-ratio restriction. These bounds give finite-size accuracy guarantees
and control the exponentially small fraction of unperturbed ground states
that survive the selection.
\end{abstract}
\maketitle

\section{Introduction}
An antiferromagnetic Ising triangle cannot make all three bonds
antiparallel. In the zero-field triangular-lattice model with only a uniform
nearest-neighbor coupling $J_1>0$, every ground-state triangle has one
parallel-spin bond. The many ways of satisfying this local rule give an
extensive degeneracy and a nonzero residual entropy per spin
~\cite{wannier,houtappel}. Early studies of further-neighbor interactions
identified ordered ground states~\cite{metcalf74,tanaka75,kaburagi74}.
A weak antiferromagnetic second-neighbor interaction $J_2>0$ favors stripe
order~\cite{korshunov05}.
Uniform stripes alternate between adjacent rows in at least one lattice
direction. On an $L\times L$ triangular-lattice torus, both primitive
lattice directions have period $L$. If $L$ is odd, that alternation cannot
close. States that retain the local ground-state
rule must instead contain interfaces between different stripe orientations.

Domain-wall descriptions give an energy cost $2J_2$ per unit length
and identify networks whose domain areas can change while their total
wall length and energy remain fixed~\cite{korshunov05}. Later work linked
stripe ordering and its thermal destruction to the topology of domain-wall
configurations~\cite{smerald16}. For the odd-$L$ torus, minimizing the
second-neighbor energy additionally requires a network that closes
periodically while obeying the local triangle rule. The resulting questions
are the least possible total wall length, the complete set of states
attaining it, and the statistical weights of their orientation compositions.

Periodic boundaries also affect the unperturbed state count. Pairing the
centers of adjacent triangles across their shared equal-spin bond gives a
dimer representation, but a valid covering need not reconstruct periodic
spins. The periodic-spin
count provides the reference for quantifying the loss of degeneracy under
the second-neighbor interaction. Its logarithm, divided by the number of
spins, is the entropy per spin. Exact results already describe its approach
to the bulk limit in several regimes. Spectral solutions determine the entropy of
infinitely long systems of fixed width~\cite{nourhani,nourhani23,au}.
Dimer methods give finite-size corrections and asymptotic error control
on tori whose two
periods grow at a bounded ratio~\cite{ksw}. These results motivate a
complementary question: how accurately can one bound the entropy of any
given finite sample, including highly elongated ones? Such a guarantee
would both control the comparison with the selected states and specify
sample dimensions for a desired entropy accuracy.

In this work, we address the geometry and counting consequences of these
periodic constraints. At each fixed odd $L\geq5$, we prove a minimum total
interface length of $4L$ within the $J_2=0$ ground-state set and classify
all $L^2(L^2-1)/2$ states attaining it. The directional composition is
measured by the fractions of triangles with each equal-spin-bond direction.
The classification gives the same number of selected states in every
allowed composition. Under the uniform measure on these states, these fractions
therefore approach the uniform distribution over all nonnegative triples
summing to one. Figure~\ref{fig:j2states} shows two such states with
different proportions and equal interface energies. The extensive entropy
is removed, yet the whole-sample proportions do not concentrate at a
single composition. We also give a sufficient finite-size coupling
interval in which these constrained minimizers are the ground states of
the full Hamiltonian.

\input{fig_results.tex}

Our second result concerns the unperturbed model on fully periodic
$M\times N$ triangular lattices. Here $M$ and $N$ are the numbers of
lattice spacings along the two primitive periodic directions, so there
are $MN$ spins. The odd tori considered above have $M=N=L$. For every
$M,N\geq3$, the entropy per spin differs from the bulk residual entropy
by at most an explicit constant times $\min(M,N)^{-2}$. The guarantee does not restrict the aspect ratio and
gives sufficient sample dimensions for a prescribed accuracy. We also
bound the entropy of all honeycomb dimer coverings, including those that
cannot reconstruct periodic spins. Applying the spin bound to an odd
$L\times L$ torus gives an explicit finite-size correction to the
logarithm of the selected fraction. Its leading exponential decay already
follows from the polynomial selected count and the nonzero bulk residual
entropy.

Section~II defines the model, the two ground-state counts, and the
periodic spin reconstruction used to evaluate the unperturbed count. Section~III classifies the odd-torus minimizers and their
orientation statistics. Section~IV gives the finite-size entropy bounds,
the strip results used to derive them, and their quantitative consequences.
Section~V discusses the physical interpretation and unresolved limits.
Appendices~A--C contain the counting identities, entropy-bound proofs,
and geometric classification, respectively.

\section{Model and periodic ground-state counts}
\label{sec:model}
Sites are $(x,y)\in\Z_M\times\Z_N$, with $s_{x,y}=\pm1$ and
$M,N\geq3$. The two primitive lattice directions have periods $M$ and
$N$. In coordinates with nearest-neighbor directions $(1,0)$, $(0,1)$,
and $(1,1)$, the zero-field Hamiltonian is
\begin{align}
 \mathcal H_{1,2}(s)&=J_1\mathcal A(s)+J_2\mathcal B(s),\nonumber\\
 \mathcal A(s)&=\sum_{x,y}s_{x,y}
 (s_{x+1,y}+s_{x,y+1}+s_{x+1,y+1}),\nonumber\\
 \mathcal B(s)&=\sum_{x,y}s_{x,y}
 (s_{x+1,y-1}+s_{x+2,y+1}+s_{x+1,y+2}).
 \label{eq:j2model}
\end{align}
The dimensionless bond sums $\mathcal A(s)$ and $\mathcal B(s)$
collect the nearest- and second-neighbor spin products, respectively;
their energy contributions are $J_1\mathcal A$ and $J_2\mathcal B$.
Here $J_1>0$ and $J_2\geq0$ are the corresponding antiferromagnetic
couplings. We set $k_B=1$ and use natural logarithms throughout.
Coordinates are reduced modulo the two
periods. Each sum includes three forward bonds per site; if periodic
identification makes distinct bonds share their endpoints, their
multiplicities are retained. The second-neighbor selection studied below
uses odd $M=N=L\geq5$, for which these bonds are distinct.

We distinguish two counts. The number $G_0(M,N)$ includes all periodic
spin ground states at $J_2=0$. On an odd $L\times L$ torus, let
$\mathcal C_L=\{s:\mathcal A(s)=-L^2\}$ be this unperturbed ground-state
set. The number $G_{\rm sel}(L)$ counts the states that minimize
$\mathcal B$ within $\mathcal C_L$. Both counts include distinct
translations and both global spin signs. Section~III determines
$G_{\rm sel}$; Sec.~IV bounds $G_0$ and compares the two. The constrained
minimum is also the full-model ground-state minimum in a sufficient
coupling interval established in Sec.~\ref{sec:full-window}.

\subsection{From the local triangle rule to periodic spins}
At $J_2=0$, the energy is $J_1\mathcal A$.
Each triangle has either one or three aligned bonds. There are $2MN$ triangles,
and each bond belongs to two of them, so $J_1\mathcal A\geq-J_1MN$. Equality requires
exactly one aligned bond per triangle. Place a dimer between the centers
of the two triangles sharing each aligned bond. Each triangle center is
then paired with exactly one neighbor. Denote the set of occupied dimer
edges by $\mathsf M$; each dual vertex belongs to exactly one such edge.
A matching compatible with periodic spins exists for every size in our
range, as shown below.

To reconstruct spins from a matching, fix one spin and move along the
triangular-lattice bonds. A bond crossed by a dimer preserves the spin;
any other bond reverses it. The local triangle rule guarantees consistency
around each triangle. Along the two independent periodic directions,
let $\nu_1,\nu_2$ count the reversals on paths that wind once around
the torus. An odd count would assign opposite spins to the same site
after one circuit. Periodic reconstruction therefore requires both counts
to be even. Their parities are the
matching holonomies, and matchings with the same parity pair form a
holonomy sector.

For the algebraic formulation, identify each dimer with the original bond
it crosses. For an original bond $e$, $\mathbf1_{e\in\mathsf M}$ is
one if its crossing dual edge carries a dimer and zero otherwise.
The binary variable indicating a spin reversal is
\begin{equation}
 c_{\mathsf M}(e)=1-\mathbf1_{e\in\mathsf M}\pmod2.
\end{equation}
This edge variable is a closed $\Z_2$ cochain: its sum around each triangle
vanishes modulo two,
because each triangle has two unmatched edges. Let $L_1$ and
$L_2$ be the horizontal cycle at $y=0$ and the vertical cycle at $x=0$,
containing $M$ and $N$ bonds, respectively, and set
\begin{equation}
 h_r(\mathsf M)=\sum_{e\in L_r}c_{\mathsf M}(e)\pmod2,\qquad h(\mathsf M)=(h_1,h_2).
\end{equation}
Thus $h_1,h_2\in\{0,1\}$ record the two winding parities.
Write $D_h$ for the matching count with holonomy $h$, $\dtot=\sum_hD_h$,
with the size arguments suppressed when unambiguous. Only one of these
four matching counts contributes to $G_0$.

\begin{proposition}[Spin reconstruction]
\label{prop:lift}
A matching lifts to a periodic spin ground state if and only if $h=(0,0)$.
There are exactly two lifts, related by a global spin flip. Consequently,
\begin{equation}
 G_0(M,N)=2D_{00}(M,N).
\label{eq:lift}
\end{equation}
\end{proposition}
\begin{proof}
For adjacent sites $u,v$, write $t_u=(1-s_u)/2$. Reconstruction is
the equation $t_u+t_v=c_{\mathsf M}(uv)$ over $\Z_2$. A closed cochain on the torus is exact precisely
when both periods vanish. On this connected graph any two solutions differ
by a constant, giving the two spin flips.
\end{proof}

For example, only $21$ of the $417$ matchings on the $4\times4$ torus
have $h=(0,0)$. They give $42$ periodic spin ground states, since each
matching allows two overall spin signs. Table~\ref{tab:counts} gives
further small-system counts. Loop descriptions likewise distinguish
configurations by their winding around the periodic directions
~\cite{zhang09}.

\begin{table}[tbp]
\caption{Exact counts in the conventions of Sec.~\ref{sec:model}.
The sectors are labeled by spin holonomy, not by the unshifted slope parity.
Direct spin enumeration verifies $G_0$ for all listed sizes.}
\label{tab:counts}
\begin{ruledtabular}
\begin{tabular}{ccccccc}
$(M,N)$&$D_{00}$&$D_{01}$&$D_{10}$&$D_{11}$&$\dtot$&$G_0$\\
\hline
$(3,3)$&21&7&7&7&42&42\\
$(3,4)$&34&48&11&20&113&68\\
$(3,5)$&105&139&46&19&309&210\\
$(3,6)$&430&300&38&92&860&860\\
$(4,4)$&21&132&132&132&417&42\\
$(4,5)$&356&133&420&628&1537&712
\end{tabular}
\end{ruledtabular}
\end{table}

To calculate these counts, we describe a matching as a pairing of up and
down triangles. Up and down triangles have
vertices $\{(i,j),(i+1,j),(i+1,j+1)\}$ and
$\{(i,j),(i,j+1),(i+1,j+1)\}$, respectively. Each up triangle is paired across one of its three bonds; label that
choice $t(i,j)\in\{d,h,v\}$ by the direction of the crossed bond. A down triangle is
labeled $(i,j)$ by the first vertex in its listed vertex set. Across the
$d,h,v$ bonds of up triangle $(i,j)$, its neighbors are therefore down
triangles $(i,j)$, $(i,j-1)$, and $(i+1,j)$, respectively. The pairing gives
the permutation
\begin{equation}
 \pi(i,j)=\bigl(i+\mathbf1_{t(i,j)=v},j-\mathbf1_{t(i,j)=h}\bigr).
\label{eq:permutation}
\end{equation}
Every down triangle must be paired exactly once. Thus the permutation
must be bijective, which is equivalent to
\begin{equation}
 \mathbf1_{t(k,l)=d}+\mathbf1_{t(k-1,l)=v}+\mathbf1_{t(k,l+1)=h}=1.
\label{eq:local}
\end{equation}
Summing this identity over $k$ at fixed $l$ shows that every fixed-$j$ line
has the same number $H$ of $h$ cells. Summing over $l$ at fixed $k$ similarly
gives the same number $V$ of $v$ cells on every fixed-$i$ line.
These are also the matched-bond counts on $L_1,L_2$. The pair $(H,V)$
specifies directional occupations; we refer to it as the matching slope.
Thus
\begin{equation}
 h=(M+H,N+V)\pmod2,\qquad NH+MV\leq MN.
\label{eq:slope}
\end{equation}
The inequality counts the $h$ and $v$ cells, whose totals are $NH$ and $MV$.

A compatible matching always exists. If $M$ or $N$ is even, the corresponding
alternating stripe spin configuration attains $-J_1MN$.
If both are odd, choose one permutation cycle with successive steps
$v^{M-1}h^{N-1}vh$ and set every remaining cell to $d$. The cycle visits
$M+N$ distinct cells and closes. It has $H=V=1$, so its holonomy vanishes
by Eq.~\eqref{eq:slope}; Proposition~\ref{prop:lift} supplies a ground state.

\subsection{Four-sector counts}
For the bipartite honeycomb graph, we use the determinant formulation of
Kasteleyn's method and its extensions to graphs on the torus
~\cite{kasteleyn,cimasoni,boutillier,ksw,nashoconnor}.
To extract the periodically reconstructible matchings from the total,
we compute four determinants with different boundary signs. Their linear
combinations separate the four possible parity pairs $h$.
Let $B(\sigma,\tau)$ be the $MN\times MN$ biadjacency matrix corresponding
to Eq.~\eqref{eq:permutation}, with unit $d$ edges. A $v$ edge crossing
$i=M-1$ has weight $\sigma$, and an $h$ edge crossing $j=0$ has weight $\tau$.
The signs $\sigma,\tau=\pm1$ keep track of crossings of these two
seams. Define $\Delta(\sigma,\tau)=\det B(\sigma,\tau)$. Fourier
diagonalization gives
\begin{align}
 \Delta(\sigma,\tau)
 &=\prod_{w^M=\sigma}\prod_{z^N=\tau}(1+w+z)\nonumber\\
 &=\prod_{z^N=\tau}\bigl[(1+z)^M-(-1)^M\sigma\bigr].
\label{eq:product}
\end{align}
The product is a polynomial in the twists with integer coefficients. At
$\sigma,\tau=\pm1$ it is an integer.

For the two counts needed here, the resulting identities are as follows; Appendix~\ref{app:inversion} supplies the inversion
and the determinant-sign proof.

\begin{proposition}[Finite-torus identities]
For every $M,N\geq3$,
\begin{align}
 \dtot&=\frac12\sum_{\sigma,\tau=\pm1}|\Delta(\sigma,\tau)|,\label{eq:total}\\
 G_0&=\dtot-|\Delta(1,1)|,\label{eq:deficit}\\
 G_0=\dtot&\quad\Longleftrightarrow\quad3\mid M\ \hbox{and}\ 3\mid N.
\label{eq:resonance}
\end{align}
\end{proposition}

When both periods are divisible by three, Eq.~\eqref{eq:resonance}
gives $G_0=\dtot$. This equality includes the two spin signs per compatible
matching: $D_{00}=\dtot/2$, while the other three parity groups together
contain the remaining half. Thus equal total counts do not mean that
every matching reconstructs periodic spins. The spectral zeros enter the counting identity through one
determinant rather than through a change in the local ground-state rule.

Slope-resolved counts can also be evaluated using integer polynomial
arithmetic. Appendix~\ref{app:slope-counts} gives that formula and the
counts at the boundary of the allowed slope range as well as in its
interior.

\section{Odd-torus weak-coupling selection}
\label{sec:j2}

We now ask which periodic nearest-neighbor ground states minimize an
antiferromagnetic second-neighbor perturbation. On an odd
$L\times L$ torus, a uniform stripe pattern cannot close periodically:
the spins reverse after one circuit in an alternating direction. We find
the minimum interface length between local stripe orientations and
classify every state that attains it.

Set $M=N=L$ with odd $L\geq5$. We minimize the second-neighbor bond
sum $\mathcal B$ from Eq.~\eqref{eq:j2model} over $\mathcal C_L$.
All states in $\mathcal C_L$ have the same nearest-neighbor energy,
so the perturbation ranks them by $\mathcal B$ alone. At fixed $L$,
the equal-weight measure on these selected states is obtained by taking
$J_1/T\to\infty$ first and then $J_2/T\to\infty$.
Section~\ref{sec:full-window} gives a sufficient coupling interval in which
this constrained minimization also gives the ground states of the full
Hamiltonian.

Let $h(x,y)$, $v(x,y)$, and $d(x,y)$ indicate equal spins on the bonds
from $(x,y)$ to $(x+1,y)$, $(x,y+1)$, and $(x+1,y+1)$, respectively.
Each is one for equal endpoint spins and zero otherwise. The triangle
constraints are
\begin{align}
 h(x,y)+v(x+1,y)+d(x,y)&=1,\nonumber\\
 h(x,y+1)+v(x,y)+d(x,y)&=1.
 \label{eq:j2local}
\end{align}
The local constraints conserve the number of equal-spin bonds along each
periodic line. Denote the counts on a horizontal, vertical, and diagonal
line by $H,V,D$, respectively. The first two are those in
Eq.~\eqref{eq:slope}; the same local argument gives the diagonal count.
The total numbers of equal-spin bonds in the three directions are
$LH,LV,LD$. Thus $(H,V,D)/L$ gives their fractions and, equivalently,
the fractions of triangles with each local stripe orientation.
We call a fixed triple a slope sector. It records the directional
composition of a state. The holonomy sector of Sec.~II instead records
the two parities that determine whether a matching reconstructs periodic
spins; every state considered here already satisfies that compatibility
condition.
Periodic spins on an odd line require an odd number of equal bonds, so
the possible slope sectors obey
\begin{equation}
 H,V,D\in\{1,3,5,\ldots\},\qquad H+V+D=L.
 \label{eq:j2sectors}
\end{equation}
Figure~\ref{fig:j2states}(a),(b) shows two $L=15$ minimizers with different
directional compositions but the same energy. Panel (c) places their
compositions among all the allowed triples. The result below explains
both the equal energies and the equal weights of these triples.

\subsection{An exact minimum interface cost}
To express the second-neighbor energy as an interface cost, label each
elementary triangle by the direction of its unique equal bond. Adjacent
triangles with different labels meet at an interface, as shown in the
enlarged portion of Fig.~\ref{fig:j2states}. Let $W_H,W_V,W_D$ count the
shared edges of each direction along these interfaces. The opposite
corners of two triangles sharing an edge are second neighbors. Their spin product is
$-1+2f_e+2w_e$, where $f_e$ indicates an equal shared bond and $w_e$ a label
change. Since $\sum_e f_e=L^2$,
\begin{equation}
 \mathcal B=-L^2+2W,\qquad W=W_H+W_V+W_D.
 \label{eq:j2walls}
\end{equation}
Each interface edge therefore costs $2J_2$, recovering the wall tension
used in Ref.~\cite{korshunov05}. The remaining question is how much
interface the odd periodic boundary forces into a state.

\begin{theorem}[Odd-torus selection]
\label{thm:j2}
In every sector in Eq.~\eqref{eq:j2sectors},
\begin{align}
 W_H&\geq2(V+D),& W_V&\geq2(H+D),\nonumber\\
 W_D&\geq2(H+V),& \min W&=4L,\nonumber\\
 &&\min\mathcal B&=-L^2+8L.
 \label{eq:j2minimum}
\end{align}
At fixed $(H,V,D)$, every minimizer is obtained from one of the two
configurations in Appendix~\ref{app:j2construct} by a translation,
possibly followed by a global spin reversal. There are
exactly $4L^2$ distinct periodic spin minimizers in each sector. Their
total number is
\begin{equation}
 G_{\rm sel}(L)=\frac{L^2(L^2-1)}2.
 \label{eq:j2degeneracy}
\end{equation}
In particular, $L^{-2}\ln G_{\rm sel}(L)\to0$.
\end{theorem}

The theorem gives the same minimum second-neighbor energy in every slope
sector. Summing its three wall bounds gives $W\geq4L$, and the classified
states attain all three bounds at once. Their degeneracy grows only
polynomially, although the unperturbed manifold contains exponentially
many states.

Each orientation occupies one region that, in lattice coordinates adapted
to that orientation, is a rectangle with one corner rectangle removed.
The two families differ in which of two opposite corners is removed
from the diagonal-orientation region. At fixed composition they form two
distinct orbits under translations and global spin reversal, and spatial
inversion exchanges them up to these operations. Figure~\ref{fig:families}
compares the two families at the same composition. The insets isolate
the opposite corner cuts in the green $D$ regions. Each family contributes
$L^2$ distinct translations and two overall spin signs, giving $4L^2$
states per composition. The orbit count is established in
Appendix~\ref{app:j2orbits}; completeness is proved in
Appendix~\ref{app:j2rigidity}.

\input{fig_families.tex}

The minimum energy exceeds the ideal uniform-stripe value
$-(J_1+J_2)L^2$ by $8J_2L$. A uniform stripe is not an admissible periodic
state on this odd torus. This comparison measures the excess imposed by
the boundary condition, proportional to the linear size, rather than an
excitation gap above an allowed stripe ground state.

Below we outline the lower bound, its saturation, and the
classification of the equality case; Appendix~\ref{app:j2bound} gives the
full proof.

\begin{table}[tbp]
\caption{Minimum second-neighbor bond sum and number of selected states
within the periodic nearest-neighbor ground-state manifold $\mathcal C_L$.
The last column is obtained from an exact integer transfer trace,
independent of the classification proof, and agrees with
Eq.~\eqref{eq:j2degeneracy}.}
\label{tab:j2counts}
\begin{ruledtabular}
\begin{tabular}{rrr}
$L$ & $\min\mathcal B$ & $G_{\rm sel}(L)$\\
\hline
5 & 15 & 300\\
7 & 7 & 1176\\
9 & $-9$ & 3240\\
11 & $-33$ & 7260
\end{tabular}
\end{ruledtabular}
\end{table}

For the lower bound, read a fixed spin configuration one diagonal at a
time, using $t=x-y$ and ring coordinate $u=x$. The following particle
motion describes changes between these spatial slices. The local labels $h$, $v$, and $d$
represent, respectively, a particle that stays, a particle that moves one
site to the right, and a hole; their precise index shifts are given in
Appendix~\ref{app:j2lower}. A consecutive run of moving particles displaces
one hole to the left by less than $L$. Over a full period the total
displacement is $LV$. A cyclic shift by $k$ hole labels carries total
displacement $kL$. Periodic closure restores the set of $D$ hole positions,
but a tracked hole can end at another hole's initial position. The hole
labels therefore shift by $V$. Because the holes cannot pass one another, at least $D+V$ runs are
needed for this return (Lemma~\ref{lem:j2jump}),
each contributing two horizontal interfaces. Hence $W_H\geq2(D+V)$;
cyclic permutation gives the other two bounds and $W\geq4L$.
Appendix~\ref{app:j2lower} proves the lemma and applies it to the spin model.

For saturation, Appendix~\ref{app:j2construct} gives explicit supports
for both families at every positive odd triple. There are $(L^2-1)/8$
such triples; the $4L^2$ states per triple give the total count in
Eq.~\eqref{eq:j2degeneracy}.

The completeness step rules out other interface geometries at the same
energy. Treat each region of one local stripe orientation as a set of
paired triangles. Equality fixes its two directional run counts. A
periodic-profile inequality then forces each orientation to occupy one
connected component; the three resulting disks have six wall junctions.
The conserved diagonal counts then fix the corner-cut sizes,
giving precisely the two families described above.
Appendix~\ref{app:j2rigidity} gives the full argument.

The zero entropy density also follows from the minimum wall count
$W=4L$ alone, by the direct counting bound in
Appendix~\ref{app:j2bound}; it does not require the completeness argument.

Related ground-state ordering was studied in extended-range lattice-gas
models~\cite{kaburagi74}. Finite-cluster enumerations have resolved local motifs and spin-flip
connectivity~\cite{krawczyk11,krawczyk10}. Here independent integer-transfer
traces reproduce the finite counts in Table~\ref{tab:j2counts}.
At $L=5$, comparison with the motif census of
Ref.~\cite{krawczyk10} gives the same minimum; the traces and
motif conversion are included in the verification archive~\cite{verification}.

\subsection{Equal composition weights and orientation fluctuations}
The classification determines both the total degeneracy and the
statistical weights of the directional compositions. Each dot in
Fig.~\ref{fig:j2states}(c) represents $4L^2$ selected spin states. Under
the equal-weight measure on all constrained minimizers, every dot thus
has probability $8/(L^2-1)$. The normalized slopes $H/L$, $V/L$, and $D/L$ are the area fractions
of the three local stripe orientations. Writing
$(H_1,H_2,H_3)=(H,V,D)$ and $H_i=1+2n_i$ turns the sector measure
into the uniform distribution on nonnegative integer triples
$n_1+n_2+n_3=(L-3)/2$. In particular,
\begin{equation}
 \mathbb E\frac{H_i}{L}=\frac13,\qquad
 \operatorname{Var}\frac{H_i}{L}=\frac{1-9/L^2}{18}.
 \label{eq:j2slopevariance}
\end{equation}
As $L\to\infty$ through odd sizes, $(H,V,D)/L$ converges in distribution
to the uniform measure on the triangle of nonnegative fractions summing
to one (the unit simplex), equivalently
Dirichlet$(1,1,1)$. This follows by Riemann sums over the composition
lattice. Thus eliminating the extensive entropy does not fix the global
stripe proportions. Their nonzero limiting variance quantifies the
absence of concentration in this uniform selected-state ensemble.

The same composition fluctuations can be expressed through a height
field, an auxiliary scalar whose increments are determined by the matching. The
increments sum to zero around a small closed path, but the height can
change after going once around a periodic direction. These net changes are called height periods.

Orient dual edges from the up-triangle to the down-triangle sublattice.
For the matching $\mathsf M$ associated with the spin state, define the
auxiliary integer current
\begin{equation}
 \eta_{\mathsf M}(e)=3\mathbf1_{e\in\mathsf M}-1
\end{equation}
This current takes the values $2$ on occupied edges and $-1$ on unoccupied edges.
Its divergence vanishes because each dual vertex has one occupied and two
unoccupied edges, whose signed contributions sum to zero.
Rotating this current assigns height increments to the original lattice
bonds. Their sum around each elementary triangle is zero. Choose orientations
so that its periods around the two length-$L$ directions are
\begin{equation}
 \Phi_1=3H-L,\qquad \Phi_2=3V-L.
\label{eq:height}
\end{equation}
Then $(\Phi_1,\Phi_2)\bmod2=h(\mathsf M)$. Spin compatibility selects even height
periods. Thus the periods record the global winding of a height field
whose increments are locally consistent.
The height periods express the same fluctuation:
$\mathbb E\Phi_1=0$ and
$\operatorname{Var}(\Phi_1/L)=(1-9/L^2)/2$.
The net height change around a periodic direction therefore fluctuates
on the scale of $L$.

\subsection{Directional couplings select orientations}
Directional couplings lift the composition degeneracy left by the
isotropic interaction. The same wall bounds determine the selected slopes. Write
$H_H=H,H_V=V,H_D=D$, and let $\mathcal B_i$ be the opposite-corner sum for
shared nearest-neighbor edges of direction $i$. Thus $\mathcal B_H$ uses
displacement $(1,2)$, $\mathcal B_V$ uses $(2,1)$, and $\mathcal B_D$ uses
$(1,-1)$. Here $i\in\{H,V,D\}$ indexes the three directions, and
$K_i\geq0$ is the second-neighbor coupling multiplying $\mathcal B_i$;
the isotropic model has $K_H=K_V=K_D=J_2$. The directional identity and
lower bound give
\begin{equation}
 \mathcal B_i=-L^2+2LH_i+2W_i
 \geq-L^2+4L+2(L-2)H_i.
\end{equation}
Since the constructed states saturate all three inequalities, the exact
minimum at fixed slope is
\begin{equation}
 \min\sum_i K_i\mathcal B_i
 =(-L^2+4L)\sum_iK_i+2(L-2)\sum_iK_iH_i.
 \label{eq:j2anisotropic}
\end{equation}
It remains to minimize over the allowed slopes. Putting $H_i=1+2n_i$
with $\sum_i n_i=(L-3)/2$ shows that a minimizing sector allocates this
total among directions with the smallest
$K_i$. If there is a unique smallest $K_i$, the only minimizing sector
has $H_i=L-2$ and the other counts equal to one. If exactly two couplings share
the minimum, the count in the third direction is one. Equal couplings
leave all sectors degenerate. These statements determine the
selected sectors for nonnegative couplings; when a coupling vanishes they
do not require saturation of its unweighted wall bound.
For strictly positive couplings every weighted minimizer saturates all
three bounds. The classification then gives degeneracy $4L^2$ for a
unique least coupling, $2L^2(L-1)$ for exactly two tied least couplings, and
$L^2(L^2-1)/2$ for three equal couplings.

\subsection{A sufficient window in the full model}
\label{sec:full-window}
So far, minimization has been restricted to $\mathcal C_L$. To relate it
to the full model, compare the selected states with configurations that
pay extra nearest-neighbor energy but gain second-neighbor energy.
For any spin configuration, the second-neighbor triangle inequality gives
$\mathcal B\geq-L^2$. The two second-neighbor triangle types, with vectors
$(1,-1)$, $(1,2)$, and their sum $(2,1)$, count each such bond twice.
Let $S$ be the set of down-spin sites and $e(S)$ the number of
nearest-neighbor edges with both endpoints in $S$. The nearest-neighbor
graph is six-regular. The number $c$ of bonds joining $S$ to its
complement is therefore $c=6|S|-2e(S)$, which is even, and
$\mathcal A$ changes in multiples
of four. A state outside $\mathcal C_L$ therefore has
$\mathcal A\geq-L^2+4$. Its energy exceeds the constrained minimum
\begin{equation}
 E_* =-(J_1+J_2)L^2+8J_2L
 \label{eq:j2fullenergy}
\end{equation}
by at least $4J_1-8LJ_2$. Consequently
\begin{equation}
 0<J_2/J_1<\frac1{2L}
 \label{eq:j2window}
\end{equation}
is sufficient for the full-model ground states to coincide with the
minimizers in Theorem~\ref{thm:j2}. The second-neighbor triangle argument
uses bond incidences and remains valid when $3\mid L$, where that graph
has disconnected components. Equation~\eqref{eq:j2window} is a sufficient
finite-size interval. It compares energies at fixed $L$ and does not
locate a thermodynamic phase boundary. In particular, its decrease with
$L$ does not determine the physical stability range at fixed coupling or
justify exchanging the constrained and thermodynamic limits.

An explicit competitor also provides an upper bound on the first crossing
out of $\mathcal C_L$. Start
from the construction in sector $(1,1,L-2)$ and flip the segment
$\{(x,1):2\leq x\leq L-1\}$. This raises $\mathcal A$ by four and lowers
$\mathcal B$ by sixteen, independently of $L$; Appendix~\ref{app:j2competitor}
gives a direct bond calculation. If $r_c(L)$ is the first ratio $J_2/J_1$ at which
an outside state reaches $E_*$, then
\begin{align}
 r_c(L)&=\min_{\substack{s\notin\mathcal C_L\\\mathcal B(s)<-L^2+8L}}
 \frac{\mathcal A(s)+L^2}{-L^2+8L-\mathcal B(s)},\nonumber\\
 \frac1{2L}&\leq r_c(L)\leq\frac14.
 \label{eq:j2threshold}
\end{align}
For $J_2/J_1>1/4$ this competitor strictly beats every constrained
minimizer. Its energy equals $E_*$ at $1/4$, but this need not imply full-model
ground-state coexistence for general $L$, since another state might cross
earlier. Full-spin enumeration at $L=5$ gives $r_c(5)=1/4$, as detailed
in Appendix~\ref{app:j2competitor}; the programs and complete histogram
are included in the verification archive~\cite{verification}.
These bounds do not determine the general threshold.

\section{Entropy suppression and finite-size control}
\begin{figure*}[t]
\centering
\begin{tikzpicture}
\begin{axis}[
 name=entropyraw,width=6.65cm,height=4.40cm,scale only axis,
 font=\small,tick label style={font=\small},
 title={(a) Entropy per spin},title style={at={(0,1.04)},anchor=south west},
 xmin=2,xmax=31,xtick={3,6,12,18,24,30},ymin=0.22,ymax=0.435,
 ytick={0.24,0.28,0.32,0.36,0.40},
 xlabel={Shorter period $m$},ylabel={$s_{m,n}=\ln G_0/(mn)$},
 grid=major,grid style={gray!15},axis line style={gray!65},tick align=outside,
 legend style={at={(0.97,0.04)},anchor=south east,draw=none,
 fill=white,font=\small,inner sep=2pt,row sep=1pt},
 legend cell align=left]
\addplot[black,line width=0.8pt,no marks,forget plot] coordinates {(2,0.323065947219450514) (31,0.323065947219450514)};
\node[anchor=south east,fill=white,inner sep=1pt,font=\small] at (axis cs:30.5,0.3245) {bulk $I$};
\addplot[color=blue!75!black,line width=0.45pt,mark=*,mark size=1.55pt]
 table[x=m,y=s] {entropy_rho1.dat};
\addlegendentry{$n/m=1$}
\addplot[color=orange!90!black,line width=0.45pt,mark=square*,mark size=1.5pt]
 table[x=m,y=s] {entropy_rho4.dat};
\addlegendentry{$n/m=4$}
\addplot[color=green!50!black,line width=0.45pt,mark=triangle*,mark size=1.8pt]
 table[x=m,y=s] {entropy_rho16.dat};
\addlegendentry{$n/m=16$}
\end{axis}
% Geometric cell only: dots are sites, with no spin configuration assigned.
\node[inner sep=0pt,anchor=north east] at (entropyraw.north east) {
\begin{tikzpicture}[x=0.30cm,y=0.30cm]
 \fill[white] (-1.9,-1.0) rectangle (5.9,3.25);
 \foreach \j in {0,1,2} {
  \draw[gray!65,line width=0.35pt] ({-0.5*\j},{0.866*\j}) -- ({5-0.5*\j},{0.866*\j});
 }
 \foreach \i in {0,...,5} {
  \draw[gray!65,line width=0.35pt] (\i,0) -- ({\i-1},1.732);
 }
 \foreach \i in {0,...,4} {
  \foreach \j in {0,1} {
   \draw[gray!65,line width=0.35pt] ({\i-0.5*\j},{0.866*\j}) -- ({\i+0.5-0.5*\j},{0.866*(\j+1)});
  }
 }
 \foreach \i in {0,...,5} {
  \foreach \j in {0,1,2} {
   \fill[gray!80] ({\i-0.5*\j},{0.866*\j}) circle[radius=0.055];
  }
 }
 \draw[black,line width=0.55pt] (0,0)--(5,0)--(4,1.732)--(-1,1.732)--cycle;
 \draw[<->,line width=0.5pt] (0,-0.48)--node[below,inner sep=1pt] {$n$}(5,-0.48);
 \draw[<->,line width=0.5pt] (-0.50,-0.05)--node[left,inner sep=1pt] {$m$}(-1.50,1.682);
 \node[anchor=south,font=\small] at (2,2.0) {Periodic cell};
\end{tikzpicture}};
\begin{axis}[
 at={(8.45cm,0)},anchor=south west,width=6.65cm,height=4.40cm,scale only axis,
 font=\small,tick label style={font=\small},
 title={(b) Guaranteed accuracy},title style={at={(0,1.04)},anchor=south west},
 xmode=log,ymode=log,xmin=2.7,xmax=33,xtick={3,6,12,30},
 xticklabels={3,6,12,30},ymin=0.0001,ymax=0.8,
 ytick={0.0001,0.001,0.01,0.1},minor tick num=0,
 xlabel={Shorter period $m$},ylabel={$|s_{m,n}-I|$},
 grid=major,grid style={gray!15},axis line style={gray!65},tick align=outside,
 legend style={at={(0.98,0.98)},anchor=north east,draw=none,
 fill=white,font=\small,inner sep=2pt},legend cell align=left]
\addlegendimage{black,dashed,line width=0.8pt}
\addlegendentry{rigorous bound}
\addlegendimage{black,only marks,mark=*,mark size=1.55pt}
\addlegendentry{actual error}
\addplot[color=blue!75!black,dashed,line width=0.8pt,no marks,forget plot]
 table[x=m,y=bound] {entropy_rho1.dat};
\addplot[color=orange!90!black,dashed,line width=0.8pt,no marks,forget plot]
 table[x=m,y=bound] {entropy_rho4.dat};
\addplot[color=green!50!black,dashed,line width=0.8pt,no marks,forget plot]
 table[x=m,y=bound] {entropy_rho16.dat};
\addplot[color=blue!75!black,only marks,mark=*,mark size=1.55pt,forget plot]
 table[x=m,y=error] {entropy_rho1.dat};
\addplot[color=orange!90!black,only marks,mark=square*,mark size=1.5pt,forget plot]
 table[x=m,y=error] {entropy_rho4.dat};
\addplot[color=green!50!black,only marks,mark=triangle*,mark size=1.8pt,forget plot]
 table[x=m,y=error] {entropy_rho16.dat};
\end{axis}
\end{tikzpicture}
\caption{The shorter period controls the guaranteed accuracy of the
periodic-spin entropy on a triangular-lattice torus.
Here $m=\min(M,N)$, $n=\max(M,N)$, and $G_0$ counts periodic ground states of the nearest-neighbor model
($J_2=0$),
including both global spin flips.
(a) The entropy per spin $s_{m,n}=\ln G_0(m,n)/(mn)$ approaches the bulk value
$I=0.3230659472\ldots$ at three aspect ratios; lines join finite-product
evaluations as guides to the eye. The inset shows a periodic cell, with
opposite edges identified.
(b) The absolute deviations (symbols) lie below the rigorous bounds
$\pi\sqrt3/(3m^2)+\ln25/(mn)$ (dashed lines) in
Eq.~\eqref{eq:spinuniform}; colors and symbols identify the same aspect
ratios in both panels. The bound holds for every integer $n\geq m\geq3$,
including arbitrarily elongated tori. }
\label{fig:bound}
\end{figure*}
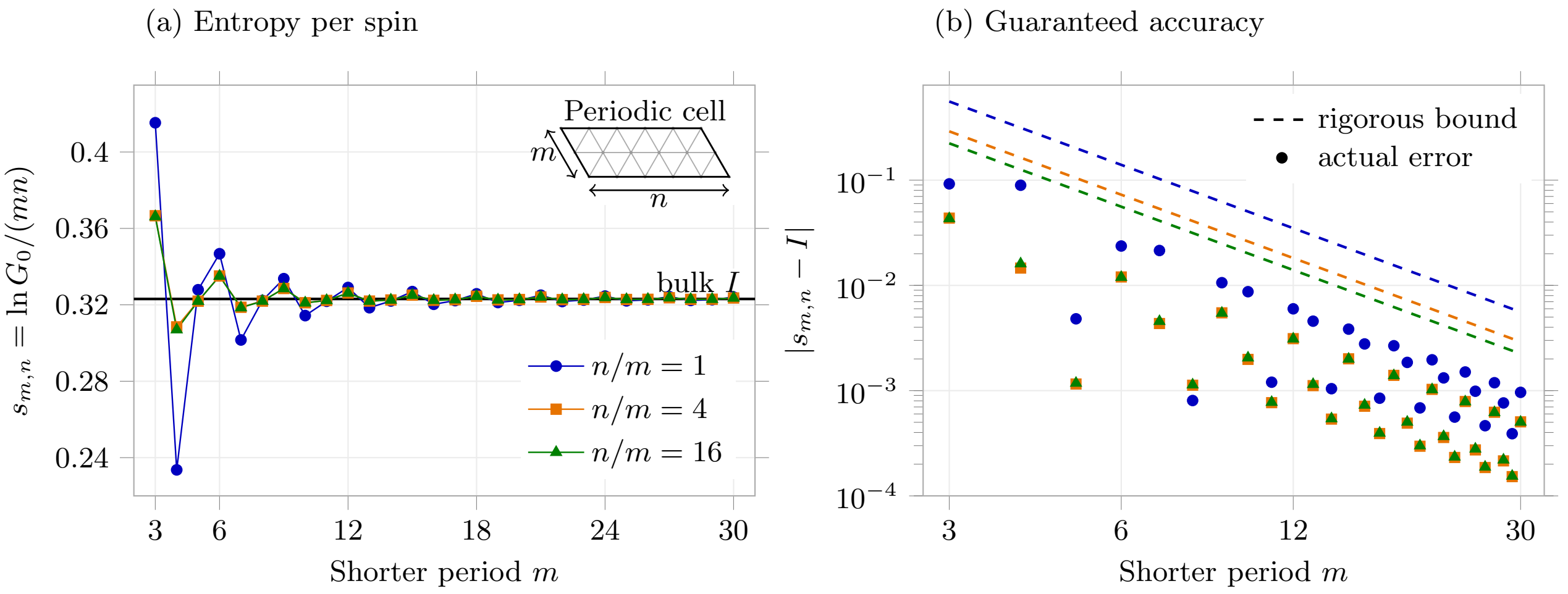

The second result concerns the nearest-neighbor model before the
second-neighbor selection. A finite periodic sample has entropy per spin
$\ln G_0(M,N)/(MN)$. We bound how far this value can be from the bulk limit,
so that the sample dimensions can be chosen for a prescribed accuracy.
Figure~\ref{fig:bound} compares the finite values with the bulk limit and
shows the guaranteed errors as the shorter period grows.
The same analysis bounds $\ln\dtot/(MN)$ for all honeycomb matchings,
including those incompatible with periodic spins. Both counts use one
triangular-lattice site, or equivalently one honeycomb unit cell, as the
normalization. Their common bulk reference is
\begin{equation}
 \begin{aligned}
 I&=\frac2\pi\int_0^{\pi/3}\ln(2\cos t)\,\dd t
   =\frac{\Cl(\pi/3)}\pi\\
  &=0.323065947219450514\ldots.
 \end{aligned}
\label{eq:anchor}
\end{equation}
The Clausen function is $\Cl(x)=\sum_{r\geq1}\sin(rx)/r^2$.

The entropy-density error is less than $5.032676/\min(M,N)^2$ for
periodic spins and $5.613369/\min(M,N)^2$ for all matchings. These bounds
hold for every $M,N\geq3$, including arbitrarily large aspect ratios, so
they allow dimensions to be chosen before evaluating a finite count.
The periodic-spin entropy is the quantity needed for the selected
fraction. We first bound the total dimer count using the positive sum
of determinant magnitudes in Eq.~\eqref{eq:total}. We then control the
subtraction in Eq.~\eqref{eq:deficit} to obtain the spin bound and apply
it to the selected fraction. Strip growth rates explain how compatibility
affects the counts; width corrections and sample-size estimates give
further quantitative consequences.

\subsection{Uniform dimer entropy bound}
\label{sec:bound}
For the full matching count, asymptotics on well-shaped tori resolve the
bounded limiting corrections to $\ln\dtot-MNI$, including arithmetic
phases~\cite{ksw}. Here we need an error estimate that also covers highly
elongated samples and finite sizes. The following bound separates a
correction controlled by the shorter period from one controlled by the
area.

\begin{theorem}[Uniform dimer bound]
\label{thm:bound}
Let $m=\min(M,N)$ and $n=\max(M,N)$, with $m\geq3$. Set
\begin{align}
 C_*={}&2\sum_{j=0}^{\infty}\left[-\ln(1-2^{-3j-1/2})
                         -\ln(1-2^{-3j-5/2})\right]\nonumber\\
     ={}&3.10642204076874649\ldots.
\end{align}
Then
\begin{align}
 \left|\frac{\ln\dtot}{MN}-I\right|
 &\leq\frac{\pi\sqrt3}{3m^2}+\frac{C_*+\ln2}{mn}\label{eq:uniform}\\
 &<\frac{5.613369}{m^2}.\nonumber
\end{align}
\end{theorem}

The two terms in Eq.~\eqref{eq:uniform} come from comparing the transverse
spectral sum with its bulk integral and controlling the finite-length
product factors, respectively. The latter are bounded by their distances from the
modes with $|1+z|=1$. Appendix~\ref{app:uniform-dimer} gives the complete
argument, including the enclosure of $C_*$.

For example, along $M=k^3$, $N=k$ with $k\geq3$, the estimate is
$\pi\sqrt3/(3k^2)+(C_*+\ln2)/k^4$, even though the aspect ratio diverges.
The unbounded aspect ratio places this sequence outside the well-shaped
hypothesis in Eq. (5) of Ref.~\cite{ksw}.

Thus the full matching entropy tends to $I$ whenever both periods grow.
On the subfamily $3\mid\gcd(M,N)$, Eq.~\eqref{eq:resonance} transfers the
same estimate to $G_0$. For a general torus, $G_0\leq\dtot$ gives only an
upper bound on the spin count. To obtain a two-sided entropy estimate,
we must also control the subtraction that selects compatible matchings.

\subsection{Uniform periodic-spin bound}
\label{sec:spinbound}
To identify a reference scale for the periodic-spin count, consider the
product in Eq.~\eqref{eq:product}. Factoring each mode by
$\max\{|1+z|^M,1\}$ gives a leading logarithmic magnitude
$M\max\{\ln|1+z|,0\}$. The remaining factors, including possible zeros
on the unit circle, are treated below. Summing the clipped logarithms on
the two transverse twist grids defines
\begin{equation}
\begin{aligned}
 \Lambda_+(N)&=\sum_{z^N=1}\max\{\ln|1+z|,0\},\\
 \Lambda_-(N)&=\sum_{z^N=-1}\max\{\ln|1+z|,0\}.
\end{aligned}
\end{equation}
A mode with $z=-1$ contributes zero to the clipped sum.

The signs $\pm$ label the grids $z^N=\pm1$ in the counting determinants.
Combining these determinants to enforce periodic spin reconstruction
leaves the rate $\Lambda_-$; the next subsection explains the
cancellation. At finite size, the corresponding exponential
$e^{n\Lambda_-(m)}$ already approximates $G_0$ within fixed factors.
Here $m=\min(M,N)$ and $n=\max(M,N)$; exchanging the two primal
coordinates preserves the graph, so $G_0(M,N)=G_0(n,m)$.

\begin{theorem}[Uniform spin bound]
\label{thm:spinbound}
For all integers $n\geq m\geq3$,
\begin{equation}
 \frac1{25}<G_0(n,m)e^{-n\Lambda_-(m)}<3.
\label{eq:spinratio}
\end{equation}
Consequently,
\begin{align}
 -\frac{\pi\sqrt3}{3m^2}-\frac{\ln25}{mn}
 &<\frac{\ln G_0}{mn}-I
 <\frac{\pi\sqrt3}{3m^2}+\frac{\ln3}{mn},\label{eq:spin-sides}\\
 \left|\frac{\ln G_0}{mn}-I\right|
 &\leq\frac{\pi\sqrt3}{3m^2}+\frac{\ln25}{mn}
 <\frac{5.032676}{m^2}.
\label{eq:spinuniform}
\end{align}
\end{theorem}
The conservative ratio bound controls the entropy loss from spin
compatibility. Its proof is computer-assisted: analytic estimates cover
$m\geq1000$, and integer intervals certify the remaining finite width range.
The programs and complete interval records are supplied in the verification archive~\cite{verification}, together with source checksums
and reproduction instructions.

The proof isolates a positive contribution from the half-shifted grid
and bounds the remaining determinant contributions. The resulting envelope
is nonincreasing in the strip length, so a width certificate extends to
every larger length. Appendix~\ref{app:spinuniform} gives the full proof
and its analytic and finite-width certification steps.

Taking logarithms gives the area correction in
Eq.~\eqref{eq:spinuniform}; the spectral rate supplies the width
correction. Together they prove the periodic-spin bulk limit whenever
$\min(M,N)\to\infty$, without arithmetic or aspect-ratio restrictions.

We can now compare the selected count from Sec.~\ref{sec:j2} with the
unperturbed periodic-spin count. For odd $L\geq5$, the uniform measure on
$\mathcal C_L$ assigns the selected states a fraction
$P_{\rm sel}(L)=G_{\rm sel}(L)/G_0(L,L)$. The numerator is known exactly; the spin
entropy bound controls the denominator. Equations~\eqref{eq:j2degeneracy} and
\eqref{eq:spinuniform} give
\begin{equation}
 \begin{aligned}
 \ln P_{\rm sel}(L)&=-IL^2+4\ln L-\ln2\\
 &\quad+\ln(1-L^{-2})+\xi_L,\\
 |\xi_L|&<5.032676.
 \end{aligned}
 \label{eq:selected-fraction}
\end{equation}
Thus $\ln P_{\rm sel}(L)=-IL^2+4\ln L+O(1)$ along odd sizes.
Isotropic selection retains every allowed odd slope sector, while its
minimizers occupy an exponentially small fraction of the full periodic-spin
nearest-neighbor ground-state ensemble. Here $P_{\rm sel}$ is a static
fraction under that uniform measure, not a cooling probability.

\subsection{Strip rates and sector suppression}
How much can the periodic-spin reconstruction condition reduce the
number of matchings? To isolate this effect, fix the width $N$ and let
the length $M$ grow. The relevant fraction is $p_{M,N}=D_{00}/\dtot$;
$J_2=0$ throughout this comparison. Define the strip entropy densities by
\begin{equation}
 \begin{aligned}
 \sD(N)&=\lim_{M\to\infty}\frac{\ln\dtot(M,N)}{MN},\\
 \sS(N)&=\lim_{M\to\infty}\frac{\ln G_0(M,N)}{MN}.
 \end{aligned}
\end{equation}

\begin{theorem}[Strip rates]
\label{thm:strip}
For fixed $N\geq3$,
\begin{equation}
 \begin{aligned}
 \sD(N)&=\frac{\max(\Lambda_+,\Lambda_-)}N,\\
 \sS(N)&=\frac{\Lambda_-}N.
 \end{aligned}
\label{eq:rates}
\end{equation}
The ordering is $\Lambda_+>\Lambda_-$ if $3\nmid N$ and
$\Lambda_+<\Lambda_-$ if $3\mid N$.
\end{theorem}

These rates determine the exponential decay of the compatible matching
fraction $p_{M,N}$. Whereas $P_{\rm sel}$ measures the reduction of the
spin ground-state count by the second-neighbor interaction, $p_{M,N}$
measures which dimer coverings represent periodic spins. Equations~\eqref{eq:lift} and
\eqref{eq:rates} imply
\begin{equation}
 \lim_{M\to\infty}\frac1M\ln p_{M,N}=-N\gamma_N,
 \qquad \gamma_N=\sD(N)-\sS(N).
\label{eq:fraction}
\end{equation}
For $3\nmid N$, this fraction decays exponentially with length. For
$3\mid N$, its exponential rate is zero; the finite-size equality of
Eq.~\eqref{eq:resonance} still also requires $3\mid M$. The gap $\gamma_N$
is an entropy-density deficit, not an excitation energy.

Table~\ref{tab:striprates} illustrates both possibilities: widths divisible
by three have equal strip rates, while other widths have a positive
entropy-density deficit.

\begin{table}[tbp]
\caption{Entropy per site in the infinite-length limit at fixed width
$N$: $\sD$ includes all matchings, $\sS$ counts periodic-spin ground
states, and $\gamma_N=\sD-\sS$ is their entropy-density difference.
Values are evaluated from the defining spectral sums.
All logarithms are natural and $k_B=1$.}
\label{tab:striprates}
\begin{ruledtabular}
\begin{tabular}{cccc}
$N$&$\sD(N)$&$\sS(N)$&$\gamma_N$\\
\hline
3&0.3662040962&0.3662040962&0\\
4&0.3465735903&0.3069867943&0.0395867960\\
5&0.3311141661&0.3218875825&0.0092265836\\
6&0.3350175129&0.3350175129&0\\
7&0.3303315732&0.3185104448&0.0118211284\\
8&0.3267801923&0.3219297619&0.0048504304
\end{tabular}
\end{ruledtabular}
\end{table}

The distinction between the rates comes from determinant cancellation.
For the full matching count, Eq.~\eqref{eq:product} assigns every mode with $|1+z|\ne1$ its clipped logarithm. Modes with $|1+z|=1$ occur only on
the $+$ transverse grid, and a suitable choice of longitudinal twist avoids
cancellation. When $3\nmid N$, the periodic-spin formula
\eqref{eq:deficit} subtracts a determinant with the dominant rate.
To track this subtraction at all widths, write
\begin{align}
 U_M&=\tfrac12\bigl(|\Delta(1,-1)|+|\Delta(-1,-1)|\bigr),\nonumber\\
 R_M&=\tfrac12\bigl(|\Delta(-1,1)|-|\Delta(1,1)|\bigr).
\end{align}
Then $G_0=U_M+R_M$ and $M^{-1}\ln U_M\to\Lambda_-$. For $3\nmid N$, set
$\delta_N=\min_{z^N=1}|\ln|1+z||$, omitting $z=-1$ if present.
Factoring out the leading magnitude in each determinant shows
\begin{equation}
 R_M=O\bigl(e^{M(\Lambda_+-\delta_N)}\bigr).
\end{equation}
Appendix~\ref{app:strip-order} proves
$\delta_N>\Lambda_+-\Lambda_->0$. If $3\mid N$, the simpler
$R_M=O(e^{M\Lambda_+})$ suffices since $\Lambda_+<\Lambda_-$. Thus
$R_M/U_M\to0$ in both cases: the subtraction leaves $U_M$ as the
leading spin contribution, establishing the rate $\Lambda_-/N$.

This spectral rate can also be read from the fermionic cylinder description
of Ref.~\cite{nourhani}, Eqs. (5)--(7): the mode energy is
$\epsilon(q)=-\ln(2\cos(q/2))$, and even fermion number, corresponding to
periodic spins, selects half-integer momenta. Filling the negative-energy
modes in conjugate pairs gives $\Lambda_-$.
Here $\gamma_N$ compares the periodic-spin ensemble with the full matching
ensemble; it is distinct from the excitation gaps governing mutual information
in Ref.~\cite{nourhani}.

The same half-shifted rate follows from the zero-temperature cylinder
solution of Ref.~\cite{au}; Appendix~\ref{app:strip-order} gives the
explicit correspondence. The finite determinant calculation here tracks
how periodic-spin compatibility selects this rate.

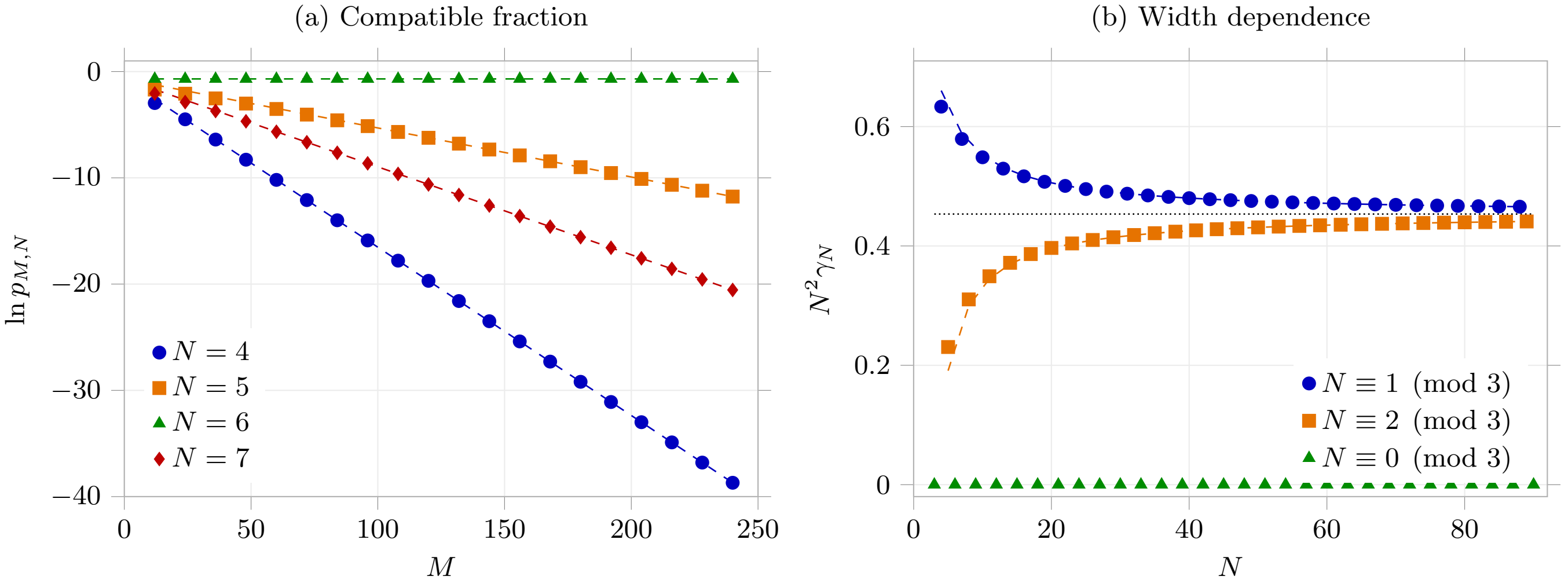
\begin{figure*}[t]
\centering
\begin{tikzpicture}
\begin{axis}[width=2.85in,height=1.96in,scale only axis,font=\small,tick label style={font=\small},legend style={font=\small,draw=none,fill=white},grid=major,grid style={gray!15},axis line style={gray!60},tick align=outside,xmin=0,xmax=250,ymin=-40,ymax=1,xlabel={$M$},ylabel={$\ln p_{M,N}$},title={(a) Compatible fraction},legend pos=south west]
\addplot[line width=0.5pt,color=blue!75!black,only marks,mark=*,mark size=2pt] coordinates {(12,-2.96339443339) (24,-4.49399237504) (36,-6.39926842789) (48,-8.29407366421) (60,-10.1940779021) (72,-12.0941529977) (84,-13.9943125659) (96,-15.894477049) (108,-17.7946430735) (120,-19.6948092443) (132,-21.5949754457) (144,-23.4951416508) (156,-25.3953078566) (168,-27.2954740624) (180,-29.1956402682) (192,-31.0958064741) (204,-32.9959726799) (216,-34.8961388858) (228,-36.7963050916) (240,-38.6964712974)};
\addlegendentry{$N=4$}
\addplot[line width=0.5pt,color=blue!75!black,dashed,no marks,forget plot] coordinates {(12,-2.5933133864) (240,-38.6964712974)};
\addplot[line width=0.5pt,color=orange!90!black,only marks,mark=square*,mark size=2pt] coordinates {(12,-1.69295119279) (24,-2.08559162835) (36,-2.52786974994) (48,-3.01118524396) (60,-3.52201887344) (72,-4.0501772061) (84,-4.58885085678) (96,-5.13376601019) (108,-5.68233683466) (120,-6.2330320796) (132,-6.78495626385) (144,-7.33758948909) (156,-7.89063117362) (168,-8.44390795335) (180,-8.99731997728) (192,-9.55080978099) (204,-10.1043443089) (216,-10.6579045511) (228,-11.211479577) (240,-11.7650631022)};
\addlegendentry{$N=5$}
\addplot[line width=0.5pt,color=orange!90!black,dashed,no marks,forget plot] coordinates {(12,-1.2467421995) (240,-11.7650475594)};
\addplot[line width=0.5pt,color=green!55!black,only marks,mark=triangle*,mark size=2pt] coordinates {(12,-0.69314718056) (24,-0.69314718056) (36,-0.69314718056) (48,-0.69314718056) (60,-0.69314718056) (72,-0.69314718056) (84,-0.69314718056) (96,-0.69314718056) (108,-0.69314718056) (120,-0.69314718056) (132,-0.69314718056) (144,-0.69314718056) (156,-0.69314718056) (168,-0.69314718056) (180,-0.69314718056) (192,-0.69314718056) (204,-0.69314718056) (216,-0.69314718056) (228,-0.69314718056) (240,-0.69314718056)};
\addlegendentry{$N=6$}
\addplot[line width=0.5pt,color=green!55!black,dashed,no marks,forget plot] coordinates {(12,-0.69314718056) (240,-0.69314718056)};
\addplot[line width=0.5pt,color=red!75!black,only marks,mark=diamond*,mark size=2pt] coordinates {(12,-2.06397154086) (24,-2.87464852456) (36,-3.71298882992) (48,-4.68205089076) (60,-5.6654332659) (72,-6.65359959723) (84,-7.64490954651) (96,-8.63730103965) (108,-9.6300523089) (120,-10.6229436521) (132,-11.6158878425) (144,-12.6088512882) (156,-13.6018218609) (168,-14.5947950864) (180,-15.5877692939) (192,-16.5807438649) (204,-17.5737185705) (216,-18.566693326) (228,-19.5596681) (240,-20.5526428809)};
\addlegendentry{$N=7$}
\addplot[line width=0.5pt,color=red!75!black,dashed,no marks,forget plot] coordinates {(12,-1.68612196546) (240,-20.5526428785)};
\end{axis}
\begin{axis}[width=2.85in,height=1.96in,scale only axis,font=\small,tick label style={font=\small},legend style={font=\small,draw=none,fill=white},grid=major,grid style={gray!15},axis line style={gray!60},tick align=outside,at={(3.55in,0)},anchor=south west,xmin=0,xmax=92,ymin=-0.02,ymax=0.71,xlabel={$N$},ylabel={$N^2\gamma_N$},title={(b) Width dependence},legend pos=south east]
\addplot[line width=0.5pt,color=blue!75!black,only marks,mark=*,mark size=2pt] coordinates {(4,0.633388735281) (7,0.579235291191) (10,0.548560502254) (13,0.52951548046) (16,0.516686440785) (19,0.507502177815) (22,0.500619729966) (25,0.495277556055) (28,0.491014211984) (31,0.487534739462) (34,0.484642189331) (37,0.482200214559) (40,0.480111513074) (43,0.478304830287) (46,0.47672681435) (49,0.47533673984) (52,0.474102989387) (55,0.473000647302) (58,0.472009816462) (61,0.47111441735) (64,0.470301315709) (67,0.469559678615) (70,0.468880492213) (73,0.4682561957) (76,0.467680400179) (79,0.467147670294) (82,0.466653352892) (85,0.466193441318) (88,0.465764466992)};
\addlegendentry{$N\equiv1\pmod3$}
\addplot[line width=0.5pt,color=blue!75!black,dashed,no marks,forget plot] coordinates {(4,0.660267767581) (7,0.588122391223) (10,0.552338071976) (13,0.531430262511) (16,0.517780151055) (19,0.508182327188) (22,0.501070286568) (25,0.495590903075) (28,0.491240698066) (31,0.487703639214) (34,0.484771439814) (37,0.482301291968) (40,0.480192031327) (43,0.478369999286) (46,0.476780295074) (49,0.475381165019) (52,0.474140291119) (55,0.47303226919) (58,0.472036854504) (61,0.471137715653) (64,0.470321532831) (67,0.469577334405) (70,0.468896001419) (73,0.468269892456) (76,0.467692556092) (79,0.467158507966) (82,0.466663056154) (85,0.466202163056) (88,0.465772335192)};
\addplot[line width=0.5pt,color=orange!90!black,only marks,mark=square*,mark size=2pt] coordinates {(5,0.230664591225) (8,0.310427546533) (11,0.349194277158) (14,0.371752048745) (17,0.386406564988) (20,0.396656396742) (23,0.404213521869) (26,0.410009313993) (29,0.414592054264) (32,0.418304793313) (35,0.421372839991) (38,0.423950133606) (41,0.426145354427) (44,0.428037378385) (47,0.429684816632) (50,0.431132118253) (53,0.432413596571) (56,0.433556157824) (59,0.434581194607) (62,0.435505927495) (65,0.436344373561) (68,0.437108057362) (71,0.437806540804) (74,0.438447823481) (77,0.439038648947) (80,0.439584741722) (83,0.440090992612) (86,0.440561605052) (89,0.441000211674)};
\addlegendentry{$N\equiv2\pmod3$}
\addplot[line width=0.5pt,color=orange!90!black,dashed,no marks,forget plot] coordinates {(5,0.191029137989) (8,0.299537564333) (11,0.344852701851) (14,0.36962268811) (17,0.385214581157) (20,0.395925195446) (23,0.403733912176) (26,0.409678251543) (29,0.414354166093) (32,0.418128228969) (35,0.421238250109) (38,0.42384522399) (41,0.426062015252) (44,0.42797008728) (47,0.429629709122) (50,0.431086425226) (53,0.43237529221) (56,0.433523732688) (59,0.434553505528) (62,0.435482095971) (65,0.436323715555) (68,0.437090033882) (71,0.437790722527) (74,0.438433865045) (77,0.439026270003) (80,0.43957371279) (83,0.440081124401) (86,0.440552740303) (89,0.440992218874)};
\addplot[line width=0.5pt,color=green!55!black,only marks,mark=triangle*,mark size=2pt] coordinates {(3,0) (6,0) (9,0) (12,0) (15,0) (18,0) (21,0) (24,0) (27,0) (30,0) (33,0) (36,0) (39,0) (42,0) (45,0) (48,0) (51,0) (54,0) (57,0) (60,0) (63,0) (66,0) (69,0) (72,0) (75,0) (78,0) (81,0) (84,0) (87,0) (90,0)};
\addlegendentry{$N\equiv0\pmod3$}
\addplot[line width=0.5pt,black,densely dotted,no marks,forget plot] coordinates {(3,0.453449841059) (90,0.453449841059)};
\end{axis}\end{tikzpicture}
\caption{Periodic-spin compatibility can suppress a matching fraction
exponentially with strip length even though its entropy-density cost
vanishes with width. (a) Natural logarithm of
$p_{M,N}=D_{00}/\dtot$, the fraction of all honeycomb matchings that lift
to periodic spins on a torus of length $M$ and width $N$. Points are
evaluated from finite products at
$M=12,24,\ldots,240$. Dashed lines are $-\ln2-MN\gamma_N$, where
$\gamma_N=\sD(N)-\sS(N)$ is the strip entropy-density deficit.
Compatibility suppresses this fraction exponentially with length for
$3\nmid N$; the displayed $N=6$ points have $p_{M,6}=1/2$ exactly.
(b) Spectral evaluations of $N^2\gamma_N$ for $3\leq N\leq90$.
Dashed curves retain terms through $N^{-4}$ in Eq.~\eqref{eq:gap};
the dotted line is $\sqrt3\pi/12$, the common limit of the two
nondivisible branches. Thus their entropy-density deficit vanishes as
$N^{-2}$, while $\gamma_N=0$ exactly for $3\mid N$.}
\label{fig:strip}
\end{figure*}

\subsection{Finite-width expansions}
At fixed width, making the torus longer approaches a strip entropy,
which generally differs from $I$. The bulk entropy is recovered by
subsequently increasing the width.
The width corrections depend on $N$ modulo three, which must be retained
in a controlled extrapolation.
Write $\chi_3(N)=0,1,-1$ when $N$ has residue $0,1,2$ modulo three,
respectively.

The first terms of the unshifted-grid rate are
\begin{equation}
 \frac{\Lambda_+}N-I=
 -\frac{\sqrt3\pi}{6N^2}+\frac{\sqrt3\pi^3}{45N^4}+O(N^{-6}),
 \quad 3\mid N,
\label{eq:firstterms}
\end{equation}
whereas for $3\nmid N$,
\begin{equation}
 \begin{aligned}
 \frac{\Lambda_+}N-I={}&\frac{\sqrt3\pi}{18N^2}
 +\frac{4\chi_3(N)\pi^2}{81N^3}\\
 &-\frac{13\sqrt3\pi^3}{1215N^4}+O(N^{-5}).
 \end{aligned}
\end{equation}
Appendix~\ref{app:expansion} gives the coefficients at arbitrary order
and an explicit remainder bound in Theorem~\ref{thm:expansion}. It uses
Fourier sampling and endpoint expansions, complementing existing exact
partitions and expansions~\cite{ivashkevich,izmailian16,bleher,izmailian}.
The low-order terms above show the main effect: finite-width corrections
depend on the width modulo three.

The same expansion determines the periodic-spin rate through the exact
sampling identity
\begin{equation}
 \Lambda_-(N)=\Lambda_+(2N)-\Lambda_+(N)
\label{eq:halfsampling}
\end{equation}
so no second endpoint calculation is needed. On the nondivisible branch this gives
\begin{equation}
 \gamma_N=\frac{\sqrt3\pi}{12N^2}
       +\frac{\chi_3(N)\pi^2}{9N^3}
       -\frac{13\sqrt3\pi^3}{648N^4}+O(N^{-5});
\label{eq:gap}
\end{equation}
on $3\mid N$, $\gamma_N=0$ exactly. For $3\nmid N$, the entropy-density
deficit therefore vanishes as the width grows, while its leading
suppression exponent per unit length in Eq.~\eqref{eq:fraction} is
$\sqrt3\pi/(12N)$.

The explicit remainder in Appendix~\ref{app:expansion} controls these
truncations without assuming convergence of the infinite series. Its
bound justifies the $O(N^{-5})$ remainder in Eq.~\eqref{eq:gap}.

\subsection{Quantitative consequences}
\label{sec:examples}
The compatible matching fraction $p_{M,N}=D_{00}/\dtot$ measures the cost of periodic
spin compatibility within the matching ensemble. Figure~\ref{fig:strip}
shows this comparison on strips. Its fraction can be tiny even when the
entropy-density difference is small: the logarithm of the fraction
accumulates over the strip area, whereas the entropy density is divided
by that area.

The dashed lines in Fig.~\ref{fig:strip}(a) use the spectral rate and the
asymptotic intercept $-\ln2$, without a fit. For $3\nmid N$, this
intercept follows by removing the leading
magnitude from every factor with $|1+z|\ne1$ in
Eq.~\eqref{eq:product}.
The remaining magnitudes tend to one. The cancellation estimate used in
Theorem~\ref{thm:strip} then gives
\begin{equation}
 p_{M,N}=\tfrac12e^{-MN\gamma_N}[1+o(1)],\qquad 3\nmid N,
\label{eq:prefactor}
\end{equation}
for fixed $N$ and $M\to\infty$. When $3\mid N$, the half-shifted branch
dominates both counts, giving $p_{M,N}\to1/2$.
The $N=6$ points in the figure have $3\mid M$ as well, so they equal
$1/2$ exactly by Eq.~\eqref{eq:resonance}.
For example, $p_{240,4}=1.5644\times10^{-17}$ whereas
$p_{240,6}=1/2$. Such large differences in compatible fractions are
consistent with both strip entropy densities approaching $I$ as
$N\to\infty$.

For uniform rejection sampling from all matchings, $p_{M,N}$ is the
acceptance probability for periodic-spin compatibility; the $240\times4$
example requires on average $1/p_{240,4}\simeq6.39\times10^{16}$
independent draws per accepted matching.

For width extrapolation, Fig.~\ref{fig:strip}(b) shows the three residue
classes separately. The $N^{-3}$ term in Eq.~\eqref{eq:gap} has opposite
signs on the two nondivisible branches. The explicit truncation envelope
in Appendix~\ref{app:expansion}, Eq.~\eqref{eq:gap-envelope}, controls the
error at finite $N$: at $N=31$, the actual error is
$1.76\times10^{-7}$, below the guaranteed $4.89\times10^{-7}$.

Finally, the uniform bounds give dimensions sufficient for a prescribed
accuracy, without a strip extrapolation. For the matching count, define
the entropy-density error $\epsilon_{M,N}=\ln\dtot/(MN)-I$ and the aspect
ratio $\rho=n/m\geq1$. Theorem~\ref{thm:bound} gives
\begin{equation}
 \begin{aligned}
 C_{\rm w}&=\pi\sqrt3/3,\qquad C_{\rm a}=C_*+\ln2,\\
 m^2|\epsilon_{M,N}|&\leq C_{\rm w}+C_{\rm a}/\rho,
 \end{aligned}
\label{eq:rho-bound}
\end{equation}
separating the width contribution $C_{\rm w}$ from the area contribution
$C_{\rm a}/\rho$. This conservative envelope guarantees accuracy before a calculation; it
does not estimate the smallest possible correction.
For a target entropy-density error $\varepsilon>0$,
a sufficient condition is
\begin{equation}
 m\geq\max\!\left\{3,\left\lceil\sqrt{(C_{\rm w}+C_{\rm a}/\rho)/\varepsilon}\right\rceil\right\}.
\label{eq:tolerance}
\end{equation}
For $\varepsilon=10^{-3}$, this condition gives $m=75$ at $\rho=1$,
$m=46$ at $\rho=16$, and $m=44$ at $\rho=64$.
These sufficient sizes can be chosen before evaluating the finite
products. For periodic spins, Eq.~\eqref{eq:spinuniform} replaces $C_{\rm a}$ by $\ln25$,
with no divisibility condition. At $\rho=16$, choosing $N=45$ and $M=720$
guarantees $|\ln G_0/(MN)-I|<10^{-3}$. The matching bound would require
a minimum dimension of 46 at this aspect ratio.
Figure~\ref{fig:bound}(b) compares this spin-entropy guarantee with the
actual errors. Increasing the long period at fixed width reduces the
area term but leaves the width term. A guaranteed approach to the bulk
entropy therefore requires the shorter period to grow as well.

We compared the finite products in Figs.~\ref{fig:bound} and
\ref{fig:strip}(a) at 80 and 120 decimal digits and evaluated the strip
rates in Fig.~\ref{fig:strip}(b) at 120 digits. An independent calculation
with exact integer determinants reproduced the spin entropy at all 84
sizes in Fig.~\ref{fig:bound} to the precision of the plotted data.
An independent spin-row calculation agrees with $2D_h$ for all four
periodic or antiperiodic seam choices at the recorded test sizes, and
small tori were checked by complete spin enumeration. The verification
archive~\cite{verification} supplies the finite-product scripts, the
figure data, and the interval certificates used in the entropy bound.

\section{Discussion}
The odd-period boundary condition frustrates the uniform stripe order
favored by the second-neighbor interaction. Within the $J_2=0$ ground-state
set, interfaces accommodate this mismatch while preserving the local
triangle rule. Their minimum cost is proportional to $L$ and vanishes per
spin. Minimizing that cost at each finite size nevertheless reduces an
exponential degeneracy to a polynomial one. The classification also fixes
the statistics of the surviving states: each allowed composition has the
same weight, giving a continuous limiting distribution of whole-sample
orientation fractions. The known freedom to change domain areas at fixed
energy does not by itself determine these weights.

The distinction between local order and global composition is essential.
Every selected state has $4L$ interface edges, and the uniform measure
includes all translations. For a fixed connected patch of triangles,
translation averaging bounds the probability of an internal interface by
a constant times $4L/L^2$. Thus a fixed patch has a single stripe
orientation with probability tending to one, even though the
whole-sample proportions have limiting variance $1/18$. For comparison, an equal
mixture of uniform stripe orientations on a compatible even-period sample
has global fractions only at the three simplex vertices, with variance
$2/9$. The Dirichlet law instead has weight throughout the simplex
interior. A fixed observation window and an area fraction measured over
the whole sample therefore probe different aspects of the limit. The
absence of concentration stated here refers to the latter observable
in the specified odd-torus ensemble.

In the antiferromagnetic regime of an odd quantum XYZ ring at zero
field, topological frustration can make the local spontaneous
magnetization decay algebraically with ring length~\cite{maric20}.
For a periodic three-leg antiferromagnetic Ising ladder of odd length
in a weak transverse field, the calculated bipartite entanglement
entropy is consistent with three delocalized quasiparticles~\cite{catalano26}.
On the odd triangular torus, the selected states accommodate the boundary
constraint through stripe interfaces while retaining one parallel-spin
bond per triangle.

The relevant counting comparison is between these selected states and all
$J_2=0$ ground states with the same periodic spin boundaries. Their ratio
has the exponential area dependence in Eq.~\eqref{eq:selected-fraction}.
A separate comparison is needed when counting spins through dimers:
only matchings that reconstruct periodic spins contribute to $G_0$.
On a strip of fixed width, their fraction of all matchings can decay
exponentially with length, even though the difference between spin and
dimer entropy densities vanishes as the width increases. Keeping the
two comparisons distinct is necessary to quantify the entropy removed
by the interaction.

The finite-size bounds give conservative accuracy guarantees for residual
entropy, including on strongly elongated tori. The strip expansions resolve
the additional dependence on width modulo three. These results complement
the asymptotics for well-shaped tori in Ref.~\cite{ksw}, the sector and
cylinder analyses of Refs.~\cite{nourhani,nourhani23}, and higher-order
toroidal dimer expansions~\cite{ivashkevich,izmailian16,bleher}.
The integer-interval certification is essential to the constants in the
periodic-spin bound; its analytic reduction is given in
Appendix~\ref{app:spinuniform} and its records in the verification
archive~\cite{verification}.

The composition limit takes weak coupling at each fixed size before
increasing odd $L$. At fixed $L$, the uniform full-model ground-state
ensemble agrees with the constrained measure throughout
$0<J_2/J_1<1/(2L)$. Its limit as $J_2/J_1\to0^+$ therefore gives the
measure studied here. The threshold satisfies
$1/(2L)\leq r_c(L)\leq1/4$; these bounds neither establish nor exclude
stability at a fixed nonzero coupling ratio as $L$ grows. That order of
limits is not determined by the present finite-size results. A further
question is how sharp the entropy bounds can be at unrestricted aspect ratio.
Boundary-dependent entropy and vortices in other domains
\cite{millane04,millane06,blakeley} raise a further comparison. Some of
those ground states have no dimer covering, so their vortex counts
cannot be directly identified with the height periods used here.

\begin{acknowledgments}
We acknowledge Gewu Intelligence Lab for providing the collaborative research
environment in which this project was developed.
GPT-assisted
symbolic, numerical, and editorial work was checked against the derivations
and exact counts reported here.
The authors take full responsibility for the scientific content and conclusions of this paper.
\end{acknowledgments}

\section*{Data availability}
The data and code supporting this article are available in the versioned
Zenodo reproducibility archive~\cite{verification}. It includes exact counting
records, integer-interval certificates, figure and table inputs, and the
programs and instructions for reproducing the computations. Analytic derivations are given in
the main text and Appendices~A--C.

\appendix
\section{Matching sectors, signs, and slope census}
The determinant formulas in Sec.~\ref{sec:model} count matchings with
signs. We give the sector inversion and finite-torus identities, then
prove the sign rule and use it to resolve counts at each slope $(H,V)$.
The holonomy sectors are recovered by grouping slopes according to
Eq.~\eqref{eq:slope}. On a general $M\times N$ torus, the height-current
construction used in Sec.~\ref{sec:j2} gives periods
$\Phi_1=3H-M$ and $\Phi_2=3V-N$. Equation~\eqref{eq:height} is their
$M=N=L$ specialization.

\subsection{Sector inversion and finite-torus identities}
\label{app:inversion}
The boundary signs in Eq.~\eqref{eq:product} distinguish the two
crossing parities. To extract an unsigned count, we also need the
permutation sign associated with each parity pair.
For $a=(a_1,a_2)\in\Z_2^2$ set
\begin{equation}
 Q(a)=(-1)^{a_1a_2+a_1[1+(N\bmod2)]+a_2[1+(M\bmod2)]}.
\label{eq:quadratic}
\end{equation}
The permutation sign of a matching with slope $(H,V)$ is $Q(H,V)$,
where arguments of $Q$ are reduced modulo two. Appendix~\ref{app:sign}
gives an elementary derivation in this convention. Thus, with
$a=h+(M,N)\pmod2$, character inversion yields
\begin{equation}
 D_h=\frac{Q(a)}4\sum_{u,v=0}^1
       (-1)^{va_1+ua_2}\Delta((-1)^u,(-1)^v).
\label{eq:inverse}
\end{equation}
This is the genus-one homology inversion of surface-dimer theory
\cite{cimasoni}, evaluated with the explicit shift in Eq.~\eqref{eq:slope}.

The finite-torus identities in Sec.~\ref{sec:model} now follow.
\begin{proof}
The only possible zero in Eq.~\eqref{eq:product} has $|z|=|1+z|=1$,
so $z=e^{\pm2\pi i/3}$. Substitution shows that only $\Delta(1,1)$ can
vanish, exactly when both dimensions are divisible by three. Nonreal roots
pair to give positive factors. The real root $z=1$, when present, gives
$2^M-(-1)^M\sigma>0$; $z=-1$ gives $-(-1)^M\sigma$. Hence a nonzero determinant
has sign
\begin{equation}
 \operatorname{sgn}\Delta(\sigma,\tau)=
 \begin{cases}
 +1,&\tau=(-1)^{N+1},\\
 -(-1)^M\sigma,&\tau=(-1)^N.
 \end{cases}
\label{eq:signrule}
\end{equation}
Substituting these signs into Eq.~\eqref{eq:inverse}, summing over $h$, and
using Eq.~\eqref{eq:lift} gives Eqs.~\eqref{eq:total} and \eqref{eq:deficit}.
The zero criterion then gives Eq.~\eqref{eq:resonance}.
\end{proof}

\subsection{The matching sign in the cell convention}
\label{app:sign}
Every nontrivial cycle of the permutation in Eq.~\eqref{eq:permutation}
consists of $v$ steps $(1,0)$ and $h$ steps $(0,-1)$. It is an embedded
essential loop: its vertices are distinct, and a lift has nonzero monotone
displacement $(Mq,-Np)$ with $p,q\geq0$. The grid embedding has no edge
crossings for $M,N\geq3$. Disjoint essential simple loops on a torus have
parallel primitive homology classes. To recall why primitivity holds, cut
along one such loop. The complement is an annulus; a transverse arc joining
its two boundary components closes to a curve of algebraic intersection
$\pm1$ with the original loop. A divisible homology class could not have
that intersection number. Parallelism follows either from the annulus
or from vanishing pairwise intersection numbers. Monotonicity fixes the
same orientation for all the cycles.

If there are $s$ nontrivial cycles, they therefore have the same primitive
pair $(p,q)$, so $H=sp$, $V=sq$, and $s=\gcd(H,V)$. Their total length is
$NH+MV$, giving permutation sign $(-1)^{NH+MV-s}$. For integers $H,V$,
$\gcd(H,V)\equiv H+V+HV\pmod2$: it is even exactly when both entries are
even. Taking $\gcd(0,0)=0$ covers the identity permutation. This proves
Eq.~\eqref{eq:quadratic} and the constant sign at each slope.

\subsection{Resolving the count by slope}
\label{app:slope-counts}
Let $\mathcal N_{M,N}(H,V)$ count matchings of slope $(H,V)$. These counts
resolve the directional occupations more finely than the four holonomy
sectors. Their signed generating function is
\begin{equation}
 \Delta(\sigma,\tau)=\sum_{H,V}\mathcal N_{M,N}(H,V)Q(H,V)\tau^H\sigma^V.
\label{eq:censusdet}
\end{equation}
The coefficients can be obtained using integer polynomial arithmetic. Put
$c=(-1)^M\sigma$, and let $\beta_i$ be the $M$ roots of $(1+x)^M=c$.
Let $e_p$ be the degree-$p$ elementary symmetric polynomial in
$\beta_0^N,\ldots,\beta_{M-1}^N$, and let $[c^V]$ denote extraction
of the coefficient of $c^V$. The power sums $q_k(c)$ and Newton recursion are
\begin{align}
 q_k(c)&=M\sum_{b=0}^{\lfloor Nk/M\rfloor}
   \binom{Nk}{bM}(-1)^{Nk-bM}c^b,\label{eq:powersum}\\
 e_0&=1,\qquad
 p e_p=\sum_{k=1}^p(-1)^{k-1}e_{p-k}q_k.
\label{eq:newton}
\end{align}
Then
\begin{equation}
 \mathcal N_{M,N}(H,V)=\bigl|[c^V]e_{M-H}(c)\bigr|.
\label{eq:census}
\end{equation}
To derive this, write $\beta_i=\omega_i-1$ with $\omega_i^M=c$.
The root-of-unity sum in $\sum_i(\omega_i-1)^{Nk}$ gives
Eq.~\eqref{eq:powersum}, and Newton's identities give Eq.~\eqref{eq:newton}.
Alternatively evaluating the resultant in Eq.~\eqref{eq:product} gives
\begin{equation}
 [\tau^H\sigma^V]\Delta
 =(-1)^{MN+H+MV}[c^V]e_{M-H}(c).
\end{equation}
Since the matching sign is constant at each slope, taking the absolute
coefficient proves Eq.~\eqref{eq:census}. All divisions by $p$ are exact
in $\Z[c]$. The arithmetic cost is $O(M^2N)$ ring multiplications; this
statement does not bound the bit cost of the growing integer coefficients.

The support lies in $NH+MV\leq MN$, and interchange of the two periods gives
\begin{equation}
 \mathcal N_{M,N}(H,V)=\mathcal N_{N,M}(V,H).
\end{equation}
On the legs, $\mathcal N(H,0)=\binom MH$ and
$\mathcal N(0,V)=\binom NV$. For the hypotenuse put $g=\gcd(M,N)$,
$M=gM'$, and $N=gN'$. Its lattice points and their counts are
\begin{equation}
 \mathcal N(tM',N'(g-t))=\binom gt,\quad 0\leq t\leq g.
\label{eq:boundary}
\end{equation}
Indeed, equality in the slope constraint excludes $d$ cells. Every step
$h$ or $v$ changes $i-j$ by $1$. Bijectivity forces cells connected by
$(i,j)\mapsto(i+1,j+1)$ to have the same type. There are $g$ such orbits,
each of length $MN/g$; choosing which $t$ orbits have type $h$ proves
Eq.~\eqref{eq:boundary}. Thus the total hypotenuse count is $2^g$.

Every integer pair $H,V\geq0$ with $H/M+V/N\leq1$ occurs, as in the
classification of periodic lozenge tilings~\cite{ibeas,dramburg}.
For a direct construction, set $F(x,y)=\lfloor xH/M+yV/N\rfloor$ on
$\Z^2$. Mark a horizontal or vertical bond when $F$ increases by one
along it, and mark a diagonal bond when $F$ is unchanged along it, using
the positive directions in Eq.~\eqref{eq:j2local}. All three increments
are zero or one. On each triangle, the horizontal and vertical increments
sum to the diagonal increment, so exactly one bond is marked. Translation by either period adds the integer $H$ or
$V$ to $F$, making the marks periodic. The marked bonds therefore give
a matching, and summing the increments along a row or column recovers
$H$ or $V$, respectively.

\section{Strip ordering, endpoint expansions, and uniform bounds}
\label{app:strip}
This appendix proves the strip ordering, endpoint expansions, and uniform
entropy bounds used in Sec.~IV. The spin count here includes
all periodic nearest-neighbor ground states, before the second-neighbor
selection of Sec.~\ref{sec:j2}. The determinant estimates control the total
matching count; a separate cancellation estimate controls the periodic-spin
count.

\subsection{Strip ordering and cancellation}
\label{app:strip-order}
Put $a=\pi/3$ and $\varphi(t)=\ln(2\cos t)$ on $[0,a]$. Let
$f(t)=\max\{\ln(2|\cos t|),0\}$, of period $\pi$, and
$\widehat f(l)=\pi^{-1}\int_{-\pi/2}^{\pi/2}f(t)e^{-2ilt}\dd t$.
Sampling at the unshifted and half-shifted grids gives
\begin{equation}
 S_N:=\Lambda_+-\Lambda_-=4N\sum_{q\geq1,\ q\ \mathrm{odd}}\widehat f(qN).
\end{equation}
Two integrations by parts, and one more for the remainder, give for integer
$l\ne0$
\begin{equation}
 \widehat f(l)=-\frac{\sqrt3\cos(2\pi l/3)}{2\pi l^2}+\epsilon_l,
 \qquad |\epsilon_l|\leq\frac{2\sqrt3+3}{4\pi|l|^3}.
\end{equation}
For the bound, $|\sin(2\pi l/3)|\leq\sqrt3/2$,
$|\varphi''(a)|=4$, and
$\int_0^a|\varphi'''(t)|\dd t=3$.
The odd cosine sum consequently yields
\begin{equation}
 S_N=\frac{B_N}{N}+E_N,\quad |E_N|\leq\frac{C_E}{N^2},
\end{equation}
where $C_E=(2\sqrt3+3)(7/8)\zeta(3)/\pi<2.2$,
$B_N=-\sqrt3\pi/4$ on $3\mid N$ and $B_N=\sqrt3\pi/12$ otherwise.
This proves $S_N<0$ on the divisible branch for $N\geq3$, and $S_N>0$
on the nondivisible branch for $N\geq5$. At $N=4$,
$S_4=\ln[4/(2+\sqrt2)]>0$.

For $3\nmid N$, every point of the unshifted transverse grid is at distance at least
$\pi/(3N)$ from a kink in the folded $t$ variable. Monotonicity of
$\ln(2\cos t)$ on each side of the kink implies
\begin{equation}
 \delta_N\geq\frac{\kappa_0}{N},\qquad \kappa_0=\frac\pi3\tan(2\pi/9).
\end{equation}
The elementary bounds $\kappa_0-\sqrt3\pi/12>0.4$ and $C_E<2.2$ show
$\delta_N>S_N$ for $N\geq6$. The remaining cases are exact:
$\delta_4=\ln\sqrt2>S_4$ and
$\delta_5=\ln\phi>\ln(2\phi^2/5)=S_5>0$, with
$\phi=(1+\sqrt5)/2$. This closes the cancellation estimate used in
Theorem~\ref{thm:strip} without assuming numerical separation of the rates.

The half-shifted branch in Eq.~\eqref{eq:rates} agrees with the
zero-temperature cylinder solution of Ref.~\cite{au}. Set all three
couplings in its convention to $-K$ and take $K\to\infty$.
The logarithm's argument is
\begin{align}
 F_K(k_1,k_2)
 &=\cosh^3(2K)-\sinh^3(2K)\nonumber\\
 &\quad+\sinh(2K)\,[\cos k_1+\cos k_2+\cos(k_1+k_2)]\nonumber\\
 &=\tfrac14e^{2K}|1+e^{ik_1}+e^{-ik_2}|^2+O(e^{-2K}).
\end{align}
Its transverse momenta are $2\pi(k+1/2)/N$. This grid has no spectral zero;
Jensen's formula then gives the residual entropy per site
$\Lambda_-/N$, in agreement with $\sS(N)$ in Eq.~\eqref{eq:rates}.
Thus the strip rate is an explicit specialization of that published formula. The determinant calculation in Sec.~IV tracks how the periodic
compatibility subtraction produces this rate.

\subsection{Endpoint expansion and its bound}
\label{app:expansion}
The bulk value also has the form $I=3\sqrt3 L(2,\chi_3)/(4\pi)$, with
$L(2,\chi_3)=\sum_{r\geq1}\chi_3(r)/r^2$. Here $\chi_3(r)$ is zero,
$1$, or $-1$ on residues $0$, $1$, or $2$ modulo three, respectively.
Put
$a=\pi/3$, $\varphi(t)=\ln(2\cos t)$,
and $A_j=\varphi^{(j)}(a)$. Define
\begin{align}
 Z_n&=\frac{(-1)^{n+1}2^{2n-1}B_{2n}}{(2n)!},\nonumber\\
 r_k&=\frac{(-1)^{k+1}2^{2k+1}B_{2k+1}(1/3)}{(2k+1)!},\nonumber\\
 b_n(N)&=\begin{cases}1,&3\mid N,\\(3^{1-2n}-1)/2,&3\nmid N,\end{cases}
\end{align}
where $B_n(x)$ are Bernoulli polynomials and $B_n=B_n(0)$.
For $k\geq0$ define
\begin{align}
 c_{2k+2}(N)&=\frac{(-1)^kA_{2k+1}b_{k+1}(N)Z_{k+1}}{2^{2k}}
                 \pi^{2k+1},\nonumber\\
 c_{2k+1}(N)&=\frac{\chi_3(N)(-1)^kA_{2k}r_k}{2^{2k}}\pi^{2k}
 \quad(k\geq1).
\label{eq:coefficients}
\end{align}

\begin{theorem}[Expansion with a remainder]
\label{thm:expansion}
For every integer $K\geq1$ and $N\geq3$,
\begin{align}
 \frac{\Lambda_+(N)}N-I
 &=\sum_{j=2}^{2K}\frac{c_j(N)}{N^j}+\mathcal R_K(N),\label{eq:expansion}\\
 |\mathcal R_K(N)|&\leq
 \frac43|A_{2K}|\frac{\zeta(2K)}{(2N)^{2K}}.
\label{eq:remainder}
\end{align}
\end{theorem}
For $l\geq1$ repeated integration by parts in
$\widehat f(l)=(2/\pi)\int_0^a\varphi(t)\cos(2lt)\dd t$ gives
\begin{align}
 \widehat f(l)=\frac2\pi\bigg[&\sum_{k=0}^{K-1}
  \frac{(-1)^kA_{2k+1}\cos(2la)}{(2l)^{2k+2}}\nonumber\\
 &+\sum_{k=1}^{K-1}
  \frac{(-1)^kA_{2k}\sin(2la)}{(2l)^{2k+1}}\bigg]+\rho_{K,l},
\end{align}
with
\begin{equation}
 |\rho_{K,l}|\leq\frac23\frac{|A_{2K}|}{(2l)^{2K}}.
\end{equation}
Indeed, $\varphi(a)=0$ and all odd derivatives vanish at zero. If
$P_0(u)=u$ and $P_{j+1}(u)=(1+u^2)P_j'(u)$, then
$\varphi^{(j)}(t)=-P_{j-1}(\tan t)$. These polynomials have nonnegative
integer coefficients, so the derivative envelope on $[0,a]$ is attained
at $a$ and equals $|A_j|$. The remaining integral is bounded by its
length times this envelope. Also $A_{2k}\in\Z$ and
$A_{2k+1}\in\sqrt3\Z$.

The Fourier coefficients are absolutely summable. Poisson sampling therefore
gives $\Lambda_+(N)/N-I=2\sum_{q\geq1}\widehat f(qN)$ exactly.
The cosine channel is
$\sum_{q\geq1}\cos(2\pi qN/3)/q^{2n}=b_n(N)\zeta(2n)$.
The sine channel is
$\sum_{q\geq1}\sin(2\pi qN/3)/q^{2k+1}
=\sqrt3\chi_3(N)L(2k+1,\chi_3)/2$.
Using $\zeta(2n)=Z_n\pi^{2n}$ and
$L(2k+1,\chi_3)=r_k\pi^{2k+1}/\sqrt3$ proves
Eq.~\eqref{eq:coefficients}. Summing $2|\rho_{K,qN}|$ proves
Eq.~\eqref{eq:remainder}. For the periodic-spin rate, Eq.~\eqref{eq:halfsampling} gives coefficient
$2^{1-j}c_j(2N)-c_j(N)$ and remainder bounded by
$2|\mathcal R_K(2N)|+|\mathcal R_K(N)|$. The bound in
Theorem~\ref{thm:expansion} has the same power as the last retained even
term. To obtain an asymptotic remainder at a prescribed order, use a
larger $K$ and bound the extra terms; $K=3$ suffices for Eq.~\eqref{eq:gap}.
On $3\mid N$ the sine channel vanishes,
so all odd-power coefficients vanish. This is a statement about the
expansion, not an assertion that the finite-width remainder is an even
function of $1/N$.

The remainder theorem provides an explicit finite-$N$ error bar for
the truncation in Eq.~\eqref{eq:gap}. Let $\gamma_N^{[4]}$ denote its three retained
terms and set
$\mathcal B_3(N)=\frac43|A_6|\zeta(6)/(2N)^6$. On $3\nmid N$,
\begin{equation}
 \begin{aligned}
 |\gamma_N-\gamma_N^{[4]}|
 \leq{}&\frac{33|c_5(N)|}{16N^5}
       +\frac{63|c_6(N)|}{32N^6}\\
       &+\frac{65}{32}\mathcal B_3(N).
 \end{aligned}
\label{eq:gap-envelope}
\end{equation}
Indeed, $\gamma_N=2[\Lambda_+(N)/N-I]-2[\Lambda_+(2N)/(2N)-I]$;
$c_5(2N)=-c_5(N)$, $c_6(2N)=c_6(N)$, and
$\mathcal B_3(2N)=\mathcal B_3(N)/64$. Equation~\eqref{eq:gap-envelope} follows by
retaining the next two coefficients in Theorem~\ref{thm:expansion}.
Here $c_5=-20\chi_3(N)\pi^4/729$ and
$c_6=2662\sqrt3\pi^5/229635$.

\begin{table}[t]
\caption{Allowed positive scaled angular distances $q_k=3md_k/(2\pi)$
from the transverse grid to $2\pi/3$ and $4\pi/3$, where $d_k$ is the
shorter circular distance to these two angles. The grid shift is
$\alpha=0$ or $1/2$. Each listed family, with $j\geq0$, has the
indicated maximum multiplicity.}
\label{tab:ladders}
\begin{ruledtabular}
\begin{tabular}{cccc}
$\alpha$&Condition&$q$&Multiplicity\\
\hline
0&$3\nmid m$&$3j+1,\ 3j+2$&2 each\\
0&$3\mid m$&$3j+3$&4\\
$1/2$&$3\nmid m$&$3j+1/2,\ 3j+5/2$&2 each\\
$1/2$&$3\mid m$&$3j+3/2$&4
\end{tabular}
\end{ruledtabular}
\end{table}

\subsection{Proof of the uniform dimer bound}
\label{app:uniform-dimer}
For Theorem~\ref{thm:bound}, we first compare each nonzero determinant
with its leading transverse-grid sum, then compare that sum with the bulk
entropy $I$. These two errors give the inverse-area and inverse-width-squared
terms in the bound. The distance multiplicities in Table~\ref{tab:ladders}
control the first error uniformly in the aspect ratio.

\begin{proof}
By symmetry, evaluate the length-$n$ product on the transverse size-$m$
grid. Write $\psi(\theta)=\max\{\ln|1+e^{i\theta}|,0\}$ and
\begin{equation}
 J_m(\alpha)=\frac1m\sum_{k=0}^{m-1}\psi\bigl(2\pi(k+\alpha)/m\bigr),
 \quad\alpha\in\{0,1/2\}.
\end{equation}
The distributional second derivative of $\psi$ has total variation
$2\sqrt3$: its smooth part has variation $\sqrt3$ and its two derivative
jumps contribute another $\sqrt3$. With Fourier convention
$\widehat\psi(l)=(2\pi)^{-1}\int_{-\pi}^{\pi}\psi(\theta)e^{-il\theta}\,\dd\theta$,
its coefficients therefore obey
$|\widehat\psi(l)|\leq\sqrt3/(\pi l^2)$ for $l\ne0$. Exact grid sampling gives
\begin{equation}
 |J_m(\alpha)-I|\leq\frac{\pi\sqrt3}{3m^2}.
\label{eq:gridbound}
\end{equation}

Put $z_k=e^{2\pi i(k+\alpha)/m}$. A mode $z_k=-1$ has a factor of unit
magnitude in Eq.~\eqref{eq:product} and contributes zero logarithmic error;
we omit it from the following estimate. For the other modes of a nonzero
determinant let $\lambda_k=\ln|1+z_k|$. After its leading magnitude is removed,
a factor with $|1+z_k|\ne1$ in Eq.~\eqref{eq:product} has logarithmic
error $e_k$ bounded by
\begin{equation}
 \ln(1-u_k)\leq e_k\leq\ln(1+u_k),\qquad u_k=e^{-n|\lambda_k|}.
\label{eq:factorbound}
\end{equation}
A mode with $|1+z_k|=1$ occurs only on the unshifted grid with $3\mid m$.
For a nonzero determinant, the corresponding product factor has magnitude
$1$, $\sqrt3$, or $2$. Its logarithmic error therefore lies between
$0$ and $\ln2$; there are at most two such factors.

Let $d_k$ be circular distance from the angle of $z_k$ to
$\{2\pi/3,4\pi/3\}$. Concavity of $\ln(2\cos(\theta/2))$ on the inner arc,
and its slope on the outer arcs, give
\begin{equation}
 |\lambda_k|\geq\frac{3\ln2}{2\pi}d_k.
\end{equation}
If $q_k=3md_k/(2\pi)$, then $u_k\leq2^{-q_k}$ since $n\geq m$.
The possible distances form the ladders in Table~\ref{tab:ladders}.
To verify their multiplicities, a point at distance $q$ from the marks
$m,2m$ modulo $3m$ must be one of $m\pm q,2m\pm q$; retaining points in
$3\alpha+3\Z$ gives the table, and coincidences decrease multiplicities.

Sum the upper and lower errors separately. For the off-unit factors define
$P=\sum\ln(1+2^{-q_k})$ and $Q=\sum-\ln(1-2^{-q_k})$, so $P\leq Q$.
The largest ladder bound is $Q\leq C_*$ on the half-shifted, nondivisible
branch. The other branches follow from geometric sums and
$-\ln(1-u)\leq u/(1-u)$: their bounds are respectively
$64/21<3.1$, $\max\{32/49,4/7+2\ln2\}<2$, and
$32u/[7(1-u)]<2.51$, where $u=1/(2\sqrt2)$ in the last branch.
The middle bound includes both unit factors. All are smaller than $C_*$.
Consequently every nonzero determinant satisfies
\begin{equation}
 \left|\frac{\ln|\Delta|}{mn}-J_m(\alpha)\right|\leq\frac{C_*}{mn}.
\end{equation}
If $R_{\max}$ is the largest normalized logarithm of the nonzero
determinants, Eq.~\eqref{eq:total} gives
$e^{mnR_{\max}}/2\leq\dtot\leq2e^{mnR_{\max}}$.
Combining this with Eq.~\eqref{eq:gridbound} proves Eq.~\eqref{eq:uniform}.
Finally $3.1<C_*<3.106422042$ and
$\pi\sqrt3/3+C_*+\ln2<5.613369$. These inequalities can be certified by
the positive series $-\ln(1-u)=\sum_{r\geq1}u^r/r$, whose tail after $L$
terms is at most $u^{L+1}/[(L+1)(1-u)]$, and geometric bounds on the
remaining ladder levels. The rational certificate supplied in the verification archive~\cite{verification} implements these enclosures.
\end{proof}

\subsection{Proof and certification of the uniform spin bound}
\label{app:spinuniform}
To prove Theorem~\ref{thm:spinbound}, we isolate a positive reference
term in the determinant expansion and bound all other terms relative to
it. This gives a two-sided comparison for the spin count despite possible
cancellations. Analytic estimates cover large widths, and interval
arithmetic covers the remaining finite range.

\begin{proof}
For $\tau=\pm1$ let
\begin{equation}
 E_\tau(n)=
 \sum_{\substack{S\subset\{z:z^m=\tau\}\\ |S|\ {\rm even}}}
       \left[\prod_{z\in S}(1+z)\right]^n.
\end{equation}
Expanding Eq.~\eqref{eq:product} and applying Eq.~\eqref{eq:inverse} gives
\begin{equation}
 G_0(n,m)=E_-(n)+(-1)^nE_+(n).
\label{eq:spin-even}
\end{equation}
Indeed, $\sum_\sigma\sigma^m\Delta(\sigma,\tau)
=2(-1)^{(n+1)m}E_\tau(n)$ and
$Q(n,m)=(-1)^{nm+n+m}$. These signs include both parities of $n$ and $m$.

On each grid, take as reference the subset
$\mathsf R_\tau=\{z:z^m=\tau,\ |1+z|>1\}$. The half-grid reference consists of conjugate
pairs, has even cardinality, and contributes the positive term
$\mathcal L_{n,m}=e^{n\Lambda_-(m)}$ to $E_-$. The full-grid reference has odd cardinality,
because it also contains $z=1$. Define
\begin{equation}
 u_z(n)=e^{-n|\ln|1+z||},
\end{equation}
with $u=0$ at $z=-1$ and $u=1$ when $|1+z|=1$. Write $u_+$
and $u_-$ for the lists on the grids $z^m=1$ and $z^m=-1$, respectively.
For either list $u$ set
\begin{equation}
 P_{\rm e,o}(u)=\frac12\left[\prod_z(1+u_z)\pm\prod_z(1-u_z)\right],
\end{equation}
where the upper sign selects even subsets. Taking symmetric differences
with $\mathsf R_\tau$ gives
\begin{align}
 |E_-/\mathcal L_{n,m}-1|&\leq P_{\rm e}(u_-)-1,\nonumber\\
 |E_+|/\mathcal L_{n,m}&\leq e^{nS_m}P_{\rm o}(u_+).
\end{align}
Here $S_m=\Lambda_+(m)-\Lambda_-(m)$. Thus, for
$F_m(n)=P_{\rm e}(u_-)+e^{nS_m}P_{\rm o}(u_+)$,
\begin{equation}
 2-F_m(n)\leq G_0(n,m)/\mathcal L_{n,m}\leq F_m(n).
\label{eq:spin-envelope}
\end{equation}
Every nonconstant half-grid term decreases with $n$. Each weighted
full-grid odd term has exponent
$n[S_m-\sum_{z\in X}|\ln|1+z||]$.
If $3\mid m$, then $S_m<0$; otherwise any nonzero odd term has excitation
energy at least $\delta_m>S_m$, as proved in Appendix~\ref{app:strip-order}.
Hence $F_m(n)$ is nonincreasing in $n$, including the unit-mode cases.

Appendix~\ref{app:spin-large} proves $F_m(m)<39/20$ for $m\geq1000$.
For $3\leq m\leq999$, the integer-interval certificate proves
$2-F_m(m)>1/25$, except at widths 4 and 7. It proves instead
$2-F_4(6)>1/25$, $2-F_7(8)>1/25$, and
$F_4(4),F_7(7)<3$. Monotonicity then leaves only the three lower-bound
cases $(n,m)=(4,4),(5,4),(7,7)$.
Their exact counts are $42$, $712$, and $2620800$.
Since $e^{\Lambda_-(4)}=2+\sqrt2<4$ and
$e^{\Lambda_-(7)}=16\cos^2(\pi/14)\cos^2(3\pi/14)<12$,
the integer inequalities $25G_0(n,m)>4^n$ or
$25G_0(n,m)>12^n$ settle these cases.
Equation~\eqref{eq:spinratio} follows.

Finally $|\Lambda_-(m)/m-I|\leq\pi\sqrt3/(3m^2)$ by
Eq.~\eqref{eq:gridbound}. Taking logarithms proves
Eqs.~\eqref{eq:spin-sides} and \eqref{eq:spinuniform}.
The rounded constant is enclosed by a positive series for $\ln25$,
Machin's formula for $\pi$, and a rational bound on $\sqrt3$.
\end{proof}

Since $F_m(n)$ is nonincreasing, it suffices to bound it at $n=m$,
apart from the explicitly handled short lengths at widths 4 and 7.

\subsubsection{Large widths}
\label{app:spin-large}
Fold transverse coordinates to $x=(k+\alpha)/m\in[-1/2,1/2)$ and put
$\phi(x)=\ln(2\cos\pi x)$. Its zeros are at $\pm1/3$.
The mode $z=-1$ has $|1+z|=0$ and activity zero, so it is omitted here.
If the index distance $d$ to a zero is at most $m/100$, then
\begin{equation}
 |\phi'(x)|\geq\pi\sqrt3-\frac{4\pi^2}{100}>5,
 \qquad m|\phi(x)|\geq5d.
\end{equation}
On the inner interval this follows from $\sec^2\theta\leq4$ up to
$\theta=\pi/3$; on the outer interval the tangent is larger.
Outside these neighborhoods monotonicity gives $|\phi|\geq1/20$.
For $m\geq1000$, the sum of all far-mode activities on either grid is
therefore bounded by
\begin{equation}
 m e^{-m/20}\leq1000e^{-50}<10^{-6}.
\end{equation}
The estimate for $S_m$ in Appendix~\ref{app:strip-order} gives
$mS_m<0.456$ if $3\nmid m$, and $mS_m<-1.35$ if $3\mid m$.

On the nondivisible branch, half-grid distances from the two zeros lie in
$j+1/6,j+5/6$, each with multiplicity at most two. Full-grid distances lie
in $j+1/3,j+2/3$, also with multiplicity at most two.
Isolate two nearest modes on each grid. Their activities and the sums of
the remaining activities obey
\begin{align}
 a&<0.435,&
 V_-&\leq\frac{2(e^{-25/6}+e^{-35/6})}{1-e^{-5}}+10^{-6}<0.038,\nonumber\\
 b&<0.189,&
 V_+&\leq\frac{2(e^{-10/3}+e^{-20/3})}{1-e^{-5}}+10^{-6}<0.075.
\end{align}
Extending near-mode ladders to infinity and adding the far-mode sum
overcounts rather than omits modes. For a list with sum $V$, its elementary
symmetric coefficient of order $j$ is at most $V^j/j!$.
The remaining even and odd partitions are consequently bounded by
$\cosh V$ and $\sinh V$. Hence
\begin{align}
 P_{\rm e}(u_-)-1
 &\leq(1+a^2)\cosh V_-+2a\sinh V_--1<0.224,\nonumber\\
 P_{\rm o}(u_+)
 &\leq(1+b^2)\sinh V_++2b\cosh V_+<0.457.
\end{align}
Since $e^{mS_m}<1.58$,
\begin{equation}
 F_m(m)-1<0.224+1.58(0.457)=0.94606<19/20.
\end{equation}

If $3\mid m$, the half-grid distances are $j+1/2$, with multiplicity
at most four. The full-grid nonunit distances are positive integers,
also with multiplicity at most four. Their activity sums satisfy
\begin{align}
 U_-&\leq\frac{4e^{-5/2}}{1-e^{-5}}+10^{-6}<0.335,\nonumber\\
 U_+&\leq\frac{4e^{-5}}{1-e^{-5}}+10^{-6}<0.028.
\end{align}
The two full-grid unit modes have activity one, so
$P_{\rm o}(u_+)=2\prod_{\rm nonunit}(1+u)$.
It follows that
\begin{align}
 P_{\rm e}(u_-)-1&<0.060,\nonumber\\
 e^{mS_m}P_{\rm o}(u_+)&<0.26\,\frac{2}{1-0.028}<0.535.
\end{align}
Thus $F_m(m)-1<0.595$ on this branch. Both branches give
$1/20<G_0/\mathcal L_{n,m}<2$ for every $n\geq m\geq1000$.

All decimal comparisons in this subsection are rationally certified.
For $x\geq0$, a positive Taylor partial sum of $e^x$ gives a lower
bound, hence an upper bound for $e^{-x}$ by reciprocation.
The degree-20 sums suffice for the near-mode enclosures, and degree 10
at $x=50$ suffices for the far tail. The elementary rational bounds
\begin{align}
 \cosh v&\leq(1-v^2/2)^{-1},&&0\leq v<\sqrt2,\nonumber\\
 \sinh v&\leq v(1-v^2/6)^{-1},&&0\leq v<\sqrt6,\nonumber\\
 e^x&\leq1+x+\frac{x^2}{2(1-x/3)},&&0\leq x<3
\end{align}
follow by bounding factorials termwise on these ranges.
The exact rational program supplied in the verification archive~\cite{verification}
checks these comparisons, the activity sums,
and the derivative and strip-gap constants.

\subsubsection{Finite widths}
For $3\leq m\leq999$, the certificate represents interval endpoints as
integers in units of $10^{-50}$. Every arithmetic operation rounds
outward. Machin's formula
$\pi=16\arctan(1/5)-4\arctan(1/239)$ is enclosed by alternating partial
sums through indices 60 and 61. At the folded spectral arguments
$0\leq x<\pi/2$, alternating cosine sums through indices 60 and 61 enclose
$2\cos x$. The tail terms decrease at these truncations.
Exact grid indices determine whether a radius is above, equal to, or
below one; no uncertain numerical classification enters the proof.
Positive interval products, reciprocals, and integer powers then enclose
$2-F_m(n)$.

The program certifies $2-F_m(m)>1/25$ at every width in this range
except 4 and 7. At those widths it certifies
$2-F_4(6)>1/25$, $2-F_7(8)>1/25$, and
$F_4(4),F_7(7)<3$. The smallest ordinary lower endpoint is at $m=n=10$:
\begin{equation}
 2-F_{10}(10)>0.04600277656683344116>1/25.
\end{equation}
The JSON supplied in the verification archive~\cite{verification}
lists all 997 width records and the additional upper checks. The three remaining short cases are the exact counts used in the
proof above; the program verifies their integer resultants with Bareiss
elimination and exact divisibility checks.
Thus the finite step is an exhaustive interval certificate over a
specified range, followed by a proved extension in length, rather than
an extrapolation from sampled sizes.

\section{Odd-torus walls, minimizers, and full-model threshold}
\label{app:j2bound}
Throughout this appendix, $L\geq5$ is odd, and a constrained minimizer
minimizes $\mathcal B$ on the nearest-neighbor ground-state manifold
$\mathcal C_L=\{s:\mathcal A(s)=-L^2\}$. The wall lower bound and
explicit constructions establish the minimum energy. The equality-case
analysis then classifies all constrained minimizers and gives their exact
count. The last two subsections compare this constrained minimum with an
outside state in the full model and compute moments of the equal-weight
measure on the slope triples $(H,V,D)$.

\subsection{Wall lower bound and an entropy consequence}
\label{app:j2lower}
The triangle constraints can be read as an exclusion dynamics on successive
diagonals. In this description, a run of moving particles displaces one
hole. The following lemma bounds the number of such runs needed for
periodic closure; it supplies the directional wall bound in
Theorem~\ref{thm:j2}.
\begin{lemma}
\label{lem:j2jump}
Consider $d\geq1$ ordered holes on a ring of integer length $L$. Each jump
moves a single hole left by a positive distance less than $L$, without
passing another hole. Suppose $C$ jumps return the unlabeled hole positions
to their initial set, with final ordered lifts $z_i'=z_{i-v}$ for $v\geq1$,
where $z_{i+d}=z_i+L$. Then $C\geq d+v$.
\end{lemma}
\begin{proof}
Every hole moves, since $z_{i-v}<z_i$. Choose the hole whose last jump
occurs first among all last jumps, and call it $i$. Write $v=qd+m$,
$0\leq m<d$. At that moment its position is $z_{i-m}-qL$.
The $m$ preceding holes $i-r$, $1\leq r\leq m$, have initial positions
at least $z_{i-m}$ and now lie strictly to its left. Each has moved more
than $qL$ and thus made at least $q+1$ jumps. For $m<r<d$, its initial
position is greater than $z_{i-m}-L$, so it has moved more than $(q-1)L$
and made at least $q$ jumps. When $q=0$ this last bound is just zero.
Hole $i$ has moved at least $qL$, with positive displacement also when
$q=0$, and therefore has made at least $q+1$ jumps. Through this moment
the total number of jumps is at least
\begin{equation}
 (q+1)+m(q+1)+(d-1-m)q=v+1.
\end{equation}
Each of the other $d-1$ holes still has its last jump to make. Hence
$C\geq v+d$, including $d=1$ and $v\geq d$.
\end{proof}

For the spin problem set $t=x-y$ and $u=x$, both modulo $L$, and define
\begin{align}
 a_t(u)&=h(u,u-t),\nonumber\\
 b_t(u)&=v(u+1,u-t),\nonumber\\
 c_t(u)&=d(u,u-t).
\end{align}
Equation~\eqref{eq:j2local} implies
\begin{align}
 a_t(u)+b_t(u)+c_t(u)&=1,\nonumber\\
 a_{t+1}(u)+b_{t+1}(u)&=a_t(u)+b_t(u-1).
 \label{eq:j2exclusion}
\end{align}
Thus $a$ denotes particles that stay and $b$ particles that move one site
to the right. The number of occupied sites is conserved. Averaging over
$t$ gives $H+V$ occupied sites and $D$ holes in each slice.
A maximal run of $b=1$ has a hole immediately to its right: a stationary
particle there would cause a collision in Eq.~\eqref{eq:j2exclusion}.
A run of length $r$ moves that hole left by $r$. Since $D>0$, $r<L$.
Distinct runs use disjoint sites, so these jumps can be performed
sequentially without passing holes.

The horizontal wall count is
\begin{align}
 W_H&=\sum_{x,y}|v(x+1,y)-v(x,y-1)|\nonumber\\
 &=\sum_{t,u}|b_t(u)-b_t(u-1)|=2C,
\end{align}
where $C$ is the number of runs over $L$ time steps. The hole positions
return to their initial set after those steps. Their total leftward
distance is $\sum_{t,u}b_t(u)=LV$. If the ordered hole indices shift by
$k$, their total displacement is
$\sum_{i=1}^D(z_i-z_{i-k})=kL$, so $k=V$. Lemma~\ref{lem:j2jump}
with $d=D,v=V$ now gives $W_H\geq2(D+V)$.
The invertible lattice map $(x,y)\mapsto(-y,x-y)$ cyclically permutes the
three undirected nearest-neighbor directions on the $L\times L$ torus. The same
argument therefore gives the other two bounds in Eq.~\eqref{eq:j2minimum}.

\subsubsection{An entropy bound without classification}
For odd $L\geq5$, the wall lower bound above and the saturating
configurations in Appendix~\ref{app:j2construct} imply that every constrained
minimizer has exactly $4L$ wall edges. This does not require the complete
classification. There are $3L^2$ possible wall edges. Fixing their set and one triangle label determines every
other label: across a nonwall edge it stays the same, and across a wall of
direction $i$ the labels must be the other two directions. There are at most
three choices for the initial label and two spin lifts. The connected
triangle adjacency graph then gives $G_{\rm sel}(L)\leq6\binom{3L^2}{4L}$. In particular,
\begin{equation}
 \frac{\ln G_{\rm sel}(L)}{L^2}
 \leq\frac{\ln6}{L^2}+\frac4L\ln\frac{3eL}{4}.
 \label{eq:j2entropy}
\end{equation}

\subsection{Explicit constructions and finite counts}
\label{app:j2construction-counts}
For every allowed directional composition $(H,V,D)$, the two families
below attain all three wall bounds. Their same-composition geometries
are shown in Fig.~\ref{fig:families}. We now give the supports and spin
recursion, verify periodic reconstruction, and count the disjoint
translation orbits. For the displayed representatives, the square-grid
$D$-region boxes are aligned by translations. With all coordinates taken
modulo 15, the spin relation in those displayed coordinates is
$s_{\rm B}(x,y)=s_{\rm A}(12-x,10-y)$.
The finite transfer calculation at the end checks the counts at the sizes
in Table~\ref{tab:j2counts}; completeness for all odd $L\geq5$ is proved
separately in Appendix~\ref{app:j2rigidity}.

\subsubsection{Explicit minimizers in every odd slope sector}
\label{app:j2construct}
For any positive odd $H,V,D$ with $H+V+D=L$, let
$I(a,m)=\{a,a+1,\ldots,a+m-1\}\pmod L$ and $I(a,0)=\varnothing$.
Let $R$ shift a set by one site in the positive $x$ direction,
$R(S)=\{x+1:x\in S\}\pmod L$. A subscript $y$ denotes the set of
$x$ positions where the corresponding bond indicator equals one in row $y$. Two choices of diagonal support are
\begin{equation}
 d_y^{\rm A}=\begin{cases}
 I(H,D),&0\leq y<H,\\
 I(H-V,V+D),&H\leq y<H+D,\\
 \varnothing,&H+D\leq y<L,
 \end{cases}
 \label{eq:j2constructA}
\end{equation}
and
\begin{equation}
 d_y^{\rm B}=\begin{cases}
 I(H,D),&0\leq y<H,\\
 \varnothing,&H\leq y<H+V,\\
 I(H,V+D),&H+V\leq y<L.
 \end{cases}
 \label{eq:j2constructB}
\end{equation}
The spin configuration associated with either bond assignment is obtained from
\begin{align}
 s(x,0)&=(-1)^{\max(0,x-H)},\qquad 0\leq x<L,\nonumber\\
 s(x,y+1)&=[2d(x-1,y)-1]s(x-1,y).
 \label{eq:j2constructspin}
\end{align}
Its initial horizontal support is $I(0,H)$. The horizontal supports for
both choices through the first block are
\begin{equation}
 h_y=I(y,H-y)\cup I(H+D,y),\qquad 0\leq y\leq H.
 \label{eq:j2constructh}
\end{equation}
The intervals are disjoint, also when the second wraps around, because
its wrapped endpoint is $y-V-1<y$. For A the remaining supports are
\begin{equation}
 h_y^{\rm A}=\begin{cases}
 I(H+D,H),&H\leq y\leq H+D,\\
 I(y,H),&H+D\leq y\leq L,
 \end{cases}
\end{equation}
and for B they are
\begin{equation}
 h_y^{\rm B}=\begin{cases}
 I(y+D,H),&H\leq y\leq H+V,\\
 I(0,H),&H+V\leq y\leq L.
 \end{cases}
\end{equation}
The formulas agree at their shared endpoints. Directly in each row block
they satisfy
\begin{equation}
 h_y\cap d_y=\varnothing,\quad d_y\subset R(h_y\cup d_y),\quad
 h_{y+1}=R(h_y\cup d_y)\setminus d_y.
 \label{eq:j2constructlocal}
\end{equation}
With $v_y=R[\Z_L\setminus(h_y\cup d_y)]$, these identities give both
triangle constraints in Eq.~\eqref{eq:j2local}. They also identify the
actual equality indicators in the spin recursion: the vertical ratio is
$2d(x-1,y)-1$ times the horizontal ratio, and taking the horizontal ratio
at the next row yields Eq.~\eqref{eq:j2constructlocal}.

\subsubsection{Closure and sector}
The initial row closes because it has $L-H=V+D$ unequal bonds, an even
number. Every diagonal $x-y=q$ contains exactly $D$ equal diagonal bonds.
To see this, the column $x=H+j$, $j=0,\ldots,D-1$, in the first $H$ rows
supplies diagonal indices $I(j+1,H)$. The remaining $D$-row block, after
indexing its rows by $j=D-1-k$, supplies $I(j+H+1,L-H)$ in both A and B.
These two intervals partition $\Z_L$ for each $j$. Hence every diagonal
contains $D$ contributions. After $L$ steps of
Eq.~\eqref{eq:j2constructspin}, the multiplier is
$(-1)^{L-D}=1$. Both spin configurations are periodic.

Each horizontal row has $H$ equal bonds. The diagonal supports have
total size $HD+D(V+D)=LD$. Vertical row sizes are $V$ in the first block,
zero in the $D$-row block, and $V+D$ in the $V$-row block, giving total
$HV+V(V+D)=LV$. The sector is therefore exactly $(H,V,D)$.

\subsubsection{Wall counts and disjoint orbits}
\label{app:j2orbits}
Let $\triangle$ denote symmetric difference of sets. The wall counts are
\begin{align}
 W_D&=\sum_y|h_y\triangle h_{y+1}|,\nonumber\\
 W_V&=\sum_y|Rh_y\triangle h_{y+1}|,\nonumber\\
 W_H&=\sum_y|R^{-1}v_y\triangle v_{y-1}|.
\end{align}
During each of the first $H$ steps, $h_y$ loses one site and gains one
distinct site. During each of the $V$ empty-diagonal steps it shifts a
proper interval by one, and during the $D$ steps it stays fixed. Thus
$W_D=2(H+V)$. In comparison with $Rh_y$, the next support differs at two
endpoints in the $H$ and $D$ blocks and agrees in the $V$ block. This gives
$W_V=2(H+D)$. Inside each nonzero vertical block $v_y$ shifts by one;
only the three block interfaces contribute to $W_H$. Their sizes are
$V,V+D,D$ for A and $D,V+D,V$ for B. Hence $W_H=2(V+D)$.
Both choices attain $W=4L$ and the minimum in Theorem~\ref{thm:j2}.

The cyclic diagonal row-width words are
$(D^H,(V+D)^D,0^V)$ for A and $(D^H,0^V,(V+D)^D)$ for B, where superscripts
give repetition counts. The three widths are distinct and each block is
nonempty, so the cyclic orders are different. Within either choice, each
directed width transition occurs once; a stabilizing translation must
therefore have zero vertical shift. A row of width $D$ is a single proper
cyclic interval and has no nonzero horizontal translation stabilizer.
Each choice consequently has $L^2$ distinct translations, and their two
orbits are disjoint. A translation sending a spin state to its negative
would stabilize its bond array and hence be the identity, which cannot
flip a spin state. Including both spin signs yields exactly $4L^2$
constructed minimizers per sector.

\subsubsection{Exact finite counts}
The independent calculation in Table~\ref{tab:j2counts} uses a transfer
state consisting of two labeled spin rows $(a,b)$. An allowed pair has
neither type of monochromatic elementary triangle between its rows.
The transition $(a,b)\to(b,c)$ has integer cost
\begin{equation}
 \sum_x\bigl[b_{x+1}c_x+b_xc_{x+2}+a_xc_{x+1}\bigr].
\end{equation}
Over a closed path of $L$ transitions the three terms count every
second-neighbor bond exactly once. Dynamic programming retains the least
cost and the number of paths attaining it. Summing the diagonal trace
over all source row pairs counts labeled torus spin arrays; no vertical
translation quotient is taken. Horizontal translations and spin inversion
reduce only that outer source sum, with each source orbit's multiplicity
restored. The calculation splits into conserved $H$ blocks and uses
overflow-checked integer additions and orbit weights. It contains no
assumed general formula for $G_{\rm sel}(L)$. The source, completed output, and
independent construction checks are supplied in the verification archive~\cite{verification}.

\subsection{Completeness of the two minimizer families}
\label{app:j2rigidity}
We prove that simultaneous equality in the three wall bounds forces the
two supports in Appendix~\ref{app:j2construct}. The four steps establish
finite phase components, use a periodic-profile
inequality to prove their uniqueness, determine the six-junction geometry,
and fix the remaining corner sizes by diagonal counts. The argument does
not use finite counts.

\subsubsection{Step 1: Finite monotone phase components}
Assume simultaneous equality in the three wall bounds. We first show that
each connected region of one stripe orientation has a finite lift, with
interval rows and columns in coordinates adapted to that orientation.
On the infinite lift the triangle equations give an integer potential
\begin{align}
 p(x+1,y)-p(x,y)&=h(x,y),\nonumber\\
 p(x,y+1)-p(x,y)&=v(x,y),\nonumber\\
 p(x+1,y+1)-p(x,y)&=1-d(x,y).
\end{align}
Its periods are $H,V$. The height normalization used in Sec.~III is
$3p(x,y)-x-y$, up to an additive constant; its periods are
$3H-L$ and $3V-L$. The affine extension of $p$ over each elementary triangle
has gradient $(1,0),(0,1),(0,0)$ in phases $H,V,D$, respectively.
Pair the two triangles sharing an equal bond. In each phase these pairs
become unit squares in the following unimodular coordinates:
\begin{center}
\begin{tabular}{cccc}
\toprule
Phase & $a$ & $b$ & Intercept $Q$\\
\midrule
$D$ & $x$ & $y$ & $p$\\
$H$ & $-x$ & $y-x$ & $p-x$\\
$V$ & $x-y$ & $-y$ & $p-y$\\
\bottomrule
\end{tabular}
\end{center}
For $H,V$ the support index is the upper-right corner of its square;
for $D$ it is the lower-left corner. The resulting common unit shift
does not affect line counts. In all three coordinate systems $Q$ is
nondecreasing in $a,b$, with directional derivatives zero or one; both
vanish precisely for the designated phase gradient. Each candidate
square is split into the two triangles sharing an edge of that direction.
Constant $Q$ therefore labels both triangles by that edge, making it
their equal bond. Conversely a paired phase cell has constant $Q$.
Continuity makes $Q$ constant on each edge-connected phase component.
For two coordinatewise comparable cells in that component, every point
of their bounding rectangle lies coordinatewise between corners with
the same $Q$ value. Monotonicity therefore makes $Q$ constant throughout
this rectangle; the preceding cell criterion puts the rectangle in the
same component.

The local wall rule also gives
\begin{align}
 W_H&=\sum|d(x,y)-d(x,y-1)|\nonumber\\
    &=\sum|v(x+1,y)-v(x,y-1)|,\nonumber\\
 W_V&=\sum|d(x,y)-d(x-1,y)|\nonumber\\
    &=\sum|h(x-1,y)-h(x,y+1)|,\nonumber\\
 W_D&=\sum|h(x,y)-h(x,y+1)|\nonumber\\
    &=\sum|v(x+1,y)-v(x,y)|,
\end{align}
with sums over all cells. No phase has a full coordinate line. For $D$,
a full horizontal or vertical line would force $H=0$ or $V=0$.
For $H$, full vertical or diagonal support would force $V=0$ or $D=0$;
for $V$, full horizontal or diagonal support would force $H=0$ or $D=0$.
All contradict Eq.~\eqref{eq:j2sectors}. Thus each cyclic variation is
twice the number of occupied runs. At equality their totals are
\begin{center}
\begin{tabular}{ccc}
\toprule
Phase & $a$-fixed lines & $b$-fixed lines\\
\midrule
$D$ & $V+D$ & $H+D$\\
$H$ & $H+V$ & $H+D$\\
$V$ & $V+D$ & $H+V$\\
\bottomrule
\end{tabular}
\end{center}
Each total is less than $L$. Since every nonempty line contributes a
run, there is an empty line in each coordinate direction. Cutting there
places every lifted phase component in a rectangle with side lengths
less than $L$. Its projection to the torus is injective. This finiteness
uses equality of the wall bounds, not height monotonicity alone.

Rectangle filling makes every row and column of a component an interval.
Connectivity makes its occupied row and column indices consecutive.
If its bounding box has cell widths $w,h$, it contributes $w+h$ runs.
Its northwest and southeast corner cells are present, since a
minimum-$a$ cell and a maximum-$b$ cell are comparable, and likewise a
minimum-$b$ cell and a maximum-$a$ cell.
The slices $t=a-b$ consequently have interval support of length
$s=w+h-1$. Their cell counts $n(t)$ start and end at one and are
$1$-Lipschitz. Indeed, if their occupied $a$ intervals on slices $t,t+1$
are $[l,r],[l',r']$, rectangle filling gives
$l'-l,r'-r\in\{0,1\}$: any larger jump or negative jump would fill a
cell beyond one of the stated endpoints. For example, $l'>l+1$ would
fill the cell $(l+1,l-t)$ on slice $t+1$. The slice counts therefore
differ by at most one.

\subsubsection{Step 2: One component per phase}
The component profiles from Step 1 must add to the conserved directional
count on every periodic line. The next lemma bounds their total support
length; together with the equality run counts, it forces one component
per phase.
\begin{lemma}
\label{lem:j2profiles}
Let $L,K$ be integers with $1\leq K<L$. Let nonzero nonnegative integer
profiles $n_i$ have finite interval supports of
lengths $s_i$, endpoint values one, and adjacent differences of absolute
value at most one. If, for every integer $t$,
\begin{equation}
 \sum_i\sum_{j\in\Z}n_i(t+jL)=K,\qquad 1\leq K<L,
 \label{eq:j2profileperiod}
\end{equation}
then $\sum_i s_i\geq L+K-1$.
\end{lemma}
\begin{proof}
Let $m(t)$ count all positive translated profiles at time $t$, counting
overlapping translates of the same profile separately. It is $L$-periodic,
$m(t)\geq1$, and $\sum_{t\bmod L}m(t)=S:=\sum_i s_i$.
Choose a longest profile A and a lift starting at $t_0$, with value one.
For each other lift B positive there write $k_B=n_B(t_0)$, so
$\sum_B k_B=K-1$. Endpoint values and the Lipschitz bound imply
$k_B\leq\lfloor(s_B+1)/2\rfloor\leq s_{\rm A}$. Thus both A and B remain
positive for $k_B$ times starting at $t_0$. Charge one overlap for B at
each of those times. At any charged time the charges are at most
$m(t)-1$, with A left uncharged. All charged times lie in fewer than
$L$ consecutive positions, since $k_B\leq K-1<L$; their residues are
distinct. Hence $S-L\geq\sum_B k_B=K-1$. For $K=1$ there are no charges,
and $m(t)\geq1$ gives the same conclusion.
\end{proof}

For a phase with sector count $K\in\{H,V,D\}$, choose one lift of each
of its $C$ torus components. Its sector lines are $a-b$ modulo $L$, so
conservation of their cell counts gives Eq.~\eqref{eq:j2profileperiod}.
The run table yields $\sum_i(w_i+h_i)=L+K$, whereas
$S=\sum_i(w_i+h_i-1)=L+K-C$.
Lemma~\ref{lem:j2profiles} therefore gives
$L+K-C\geq L+K-1$, hence $C\leq1$.
Since $K>0$, each phase has exactly one component.

\subsubsection{Step 3: Three disks and six junctions}
The finite component has no hole, since a row through a hole would have
two separated occupied intervals. It has no diagonal pinch either.
For example, cells $(-1,0),(0,-1)$ with $(0,0),(-1,-1)$ absent put the
row-0 interval wholly to the left of the row-$-1$ interval. These
intervals share no $a$ coordinate, so no component edge crosses between
the two rows. Any edge-connected path between the stated cells would
have to cross there, a contradiction. Reflection covers the other
diagonal pinch. At every boundary vertex the remaining cell boundary
then has degree two, so it consists of simple cycles. Connectivity
gives one outer cycle, and any additional cycle would enclose a hole.
Thus each phase is an injectively embedded disk.

The local possibilities follow from the six neighbors of a spin: its
same-spin neighbors form an independent set of their six-cycle.
If their number is $k$, the wall degree is six at $k=0$, three at
$k=1,3$, zero for opposite pairs at $k=2$, and two collinear edges for
the other pairs at $k=2$. At $k=0$ each phase has two diagonally touching
cells with both intervening cells absent, contradicting its single
component without pinches. This excludes degree six. At $k=1$ one
phase has a reflex corner and the other two convex corners; at $k=3$
all three are convex. Angles are measured in each phase's square
coordinates. These cases exhaust the 18 independent sets of a six-cycle.

The wall graph bounds three disk faces and is connected: a disk's
attaching circle cannot join two graph components to form a connected
surface. With $U$ degree-two vertices and $T$ junctions, its edge count
is $U+3T/2$. The torus Euler relation gives
$U+T-(U+3T/2)+3=0$, hence $T=6$.
If $N_k$ counts sites with $k$ equal incident bonds, double counting the
$L^2$ equal bonds gives $\sum_k kN_k=2L^2=2\sum_kN_k$, or
$N_3=N_1+2N_0$. Since $N_0=0$ and $N_1+N_3=T=6$, we have
$N_1=N_3=3$. The local corner rule therefore leaves exactly three
reflex corners among all phase boundaries.

The phase bounding boxes have widths and heights
$(V+D,H+D)$, $(H+V,H+D)$, and $(V+D,H+V)$ for $D,H,V$.
The phase areas are $LD,LH,LV$, so their respective deficits from the box
areas are $HV,VD,HD$, all positive. For each simple orthogonal boundary,
the total turn gives $c-r=4$, where $c,r$ count convex and reflex
corners. Thus a nonrectangular phase has $r\geq1$. The three available
reflex corners give exactly one per phase, hence five convex corners
and six sides per boundary. Tracing these six axis-aligned sides gives
a bounding rectangle with a single corner rectangle removed. For
phase $D$, rectangle filling keeps the northwest and southeast
corners, so only the southwest or northeast corner can be removed.

\subsubsection{Step 4: Diagonal counts fix the shape}
\label{app:j2shape}
For the southwest case, let $w,h$ denote the box width and height
in square-grid cells, and put the $D$ box at
$0\leq x<w=V+D$, $0\leq y<h=H+D$, and remove
$0\leq x<a$, $0\leq y<b$, with $0<a<w$, $0<b<h$, and $ab=HV$.
The polynomial $F(z)=\sum_{(x,y)\in\text{$D$ region}}z^{x-y}$
counts cells on each diagonal modulo $L$. Since every diagonal has $D$
cells, $F(z)=D(1+z+\cdots+z^{L-1})$ in
$\Z[z,z^{-1}]/(z^L-1)$. Multiplying by
$(1-z)(1-z^{-1})$ and subtracting the corner rectangle yields
\begin{equation}
 z^a+z^{-b}+z^{V-H}=z^{-H}+z^V+z^{a-b}.
\end{equation}
The residue monomials are a basis, so these exponent multisets agree
modulo $L$, including multiplicities. The exponent $a$ cannot match
$-H$, since $0<a<L-H$, or $a-b$, since $0<b<L$. Thus $a=V$.
After canceling $z^V$, $-b$ cannot match $V-b$, so $b=H$.
The northeast case follows by inversion. These are precisely the two
diagonal supports of Appendix~\ref{app:j2construct}, up to translation.

Finally, the triangle equations determine $h,v$ from this $d$.
Two choices of $h$ with the same $d$ differ by
$\delta h(x,y+1)=\delta h(x-1,y)$, constant on each $x-y$ diagonal.
Every such diagonal has a $d=1$ cell, where both choices of $h$ must
vanish. Hence $\delta h=0$, and $v$ is forced as well. Connected spin
reconstruction leaves exactly the two global signs. The disjoint-orbit
count in Appendix~\ref{app:j2orbits} therefore exhausts all minimizers.

\subsection{A full-model competing configuration}
\label{app:j2competitor}
We now compare an explicit state outside $\mathcal C_L$ with the energy
$E_*$ in Eq.~\eqref{eq:j2fullenergy}, the energy of an actual constrained
minimizer. The construction A in sector $(H,V,D)=(1,1,L-2)$ has the simple spin
formula $u(0,0)=1$ and $u(x,y)=(-1)^{x+y-1}$ at all other sites,
using representatives $0\leq x,y<L$. It has
$\mathcal A(u)=-L^2$ and $\mathcal B(u)=-L^2+8L$ by
Appendix~\ref{app:j2construction-counts}.
Flip its spins on
$F=\{(x,1):2\leq x\leq L-1\}$ to obtain $u^F$.
For any bond direction, its bond-sum change is minus twice the sum
of $u_i u_j$ over bonds with exactly one endpoint in $F$.
Counting forward and backward neighbors of each site in $F$ counts
each crossing bond once. Their sums and changes are
\begin{center}
\begin{tabular}{crr}
\toprule
Displacement & Crossing sum & Change\\
\midrule
$(1,0)$ & $0$ & $0$\\
$(0,1)$ & $-2L+4$ & $4L-8$\\
$(1,1)$ & $2L-6$ & $-4L+12$\\
$(1,-1)$ & $2L-4$ & $-4L+8$\\
$(2,1)$ & $-2L+10$ & $4L-20$\\
$(1,2)$ & $2$ & $-4$\\
\bottomrule
\end{tabular}
\end{center}
These include all periodic wraps. For example, the $(2,1)$ forward
neighbors of $x=L-2,L-1$ and the backward neighbor of $x=2$ have product
$+1$; all other products in that direction are $-1$. The latter backward
neighbor is the exceptional spin at $(0,0)$. In the $(1,2)$ direction
the forward wrap contributes $+1$, the other $L-3$ forward products are
$-1$, and all $L-2$ backward products to row $L-1$ are $+1$.
The first three changes sum to four and the last three to minus sixteen.
Thus
\begin{equation}
 \mathcal H_{1,2}(u^F)-E_*=4J_1-16J_2,
\end{equation}
which proves the upper bound in Eq.~\eqref{eq:j2threshold} for every
odd $L\geq5$.

For $L=5$ an exhaustive full-spin calculation fixes one spin and visits
all $2^{24}$ remaining arrays. A Gray-code calculation and an independent
five-row enumeration give the same complete joint histogram of
$(\mathcal A,\mathcal B)$; the programs and complete histogram are
supplied in the verification archive~\cite{verification}. Its minimum of
$4\mathcal A+\mathcal B$ is $-85=-5L^2+8L$. For any outside state with
$\mathcal B<15$, this implies
$\mathcal A+25\geq(15-\mathcal B)/4$. Together with $u^F$ it proves
$r_c(5)=1/4$. No corresponding all-size inequality is assumed here.

\subsection{Moments of the sector measure}
\label{app:j2moments}
The exact count $4L^2$ per slope triple makes the constrained-minimizer
measure uniform over the positive odd compositions of $L$.
For $m=(L-3)/2$, the uniform compositions $n_1+n_2+n_3=m$ give
$\mathbb E n_i=m/3$ and
$\mathbb E[n_i(n_i-1)]=m(m-1)/6$ by summing the composition counts.
Thus $\operatorname{Var}n_i=m(m+3)/18$. With $H_i=1+2n_i$ this is
Eq.~\eqref{eq:j2slopevariance}. The covariances of different normalized
slopes are $-(1-9/L^2)/36$, since their sum is fixed at one.

\end{document}